\documentclass[useAMS,usenatbib]{mn2e}
\pdfoutput=1
\usepackage{times}  
\usepackage{listings}
\usepackage{graphics}
\usepackage{subfigure}
\usepackage{graphicx}
\usepackage{amssymb}
\usepackage{amsmath}
\usepackage[draft]{hyperref}
\usepackage{aas_macros}
\usepackage{verbatim}
\usepackage{color}
\usepackage{url}
\usepackage[dvipsnames]{xcolor}

\long\def\symbolfootnote[#1]#2{\begingroup%
\def\thefootnote{\fnsymbol{footnote}}\footnote[#1]{#2}\endgroup}

\newcommand{\HADES}{{\tt HADES}}
\newcommand{\HEALPIX}{{\tt HEALPix}}
\newcommand{\GALFORM}{{\tt GALFORM}}

\newcommand{\Mpc}{{\,\rm Mpc}}
\newcommand{\hMpc}{\,h^{-1}{\rm Mpc}}

\newcommand{\Msol}{{\rm M_{\odot}}}

\newcommand{\band}[1]{{\rm #1}}

\input{epsf} \title{Halo detection via large-scale Bayesian inference}

\author[Merson {\it et al.}]
       {\parbox[h]{\textwidth}{Alexander~I.~Merson$^{1,2}$\thanks{E-mail:
       alex.i.merson@jpl.nasa.gov}, Jens~Jasche$^3$,
       Filipe~B.~Abdalla$^{1,4}$, Ofer~Lahav$^1$,\\
       Benjamin~Wandelt$^{5,6,7}$, D.~Heath~Jones$^8$,
       Matthew~Colless$^9$}
  \vspace*{3pt}\\
  \noindent$^1$Department of Physics and Astronomy, University College
  London, Gower Street, London WC1E 6BT\\
  $^2$Jet Propulsion Laboratory, 4800 Oak Grove Drive, Pasadena, CA 91109, USA\\
  $^3$Excellence Cluster Universe, Technische Universit\"at M\"unchen, Boltzmannstrasse 2,
  85748 Garching, Germany\\
  $^4$Department of Physics and Electronics,
  Rhodes University, PO Box 94, Grahamstown, 6140, South
  Africa\\
  $^5$Institut d'Astrophysique de Paris (IAP), UMR 7095, CNRS
  - UPMC Universit\'e Paris 06, 98bis boulevard Arago, F-75014 Paris,
  France\\
  $^6$Institut Lagrange de Paris (ILP), Sorbonne
  Universit\'es, 98bis boulevard Arago, F-75014 Paris, France\\
  $^7$ Departments of Physics and Astronomy, University of Illinois at
  Urbana-Champaign, Urbana, IL 61801, USA\\
  $^8$Department of Physics
  and Astronomy, Macquarie University, NSW, 2109,
  Australia\\
  $^9$Research School of Astronomy \& Astrophysics, The
  Australian National University, Canberra, ACT 2611, Australia\\}

\date{}

\pagerange{\pageref{firstpage}--\pageref{lastpage}}
\pubyear{\the\year}

\begin{document}

\maketitle
\title{Halo detection via Bayesian inference}
\label{firstpage}

\begin{abstract}
We present a proof-of-concept of a novel and fully
Bayesian methodology designed to detect halos of different masses in
cosmological observations subject to noise and systematic
uncertainties. Our methodology combines the previously published
Bayesian large-scale structure inference algorithm, \HADES{}, and a
Bayesian chain rule (the Blackwell-Rao Estimator), which we use to
connect the inferred density field to the properties of dark matter
halos. To demonstrate the capability of our approach we construct a
realistic galaxy mock catalogue emulating the wide-area 6-degree Field
Galaxy Survey, which has a median redshift of approximately
0.05. Application of \HADES{} to the catalogue provides us with
accurately inferred three-dimensional density fields and corresponding
quantification of uncertainties inherent to any cosmological
observation. We then use a cosmological simulation to relate the
amplitude of the density field to the probability of detecting a halo
with mass above a specified threshold. With this information we can sum
over the \HADES{} density field realisations to construct maps of
detection probabilities and demonstrate the validity of this approach
within our mock scenario. We find that the probability of successful
of detection of halos in the mock catalogue increases as a function of
the signal-to-noise of the local galaxy observations. Our proposed
methodology can easily be extended to account for more complex
scientific questions and is a promising novel tool to analyse the
cosmic large-scale structure in observations.
\end{abstract}

\begin{keywords}
methods: numerical -- methods: statistical -- galaxies: haloes -- galaxies: clusters: general -- cosmology: dark matter -- cosmology: large-scale structure of Universe
\end{keywords}


\section{Introduction}
\label{sec:intro}

The dual role of galaxy clusters, both as cosmological probes and as
unique sites for studying extreme environments of galaxy formation,
make them essential targets for next generation cosmological galaxy
surveys (e.g. see \citealt{Borgani01a, Borgani01b, Rosati02, Voit05,
Allen11} and \citealt{Kravtsov12}). Ongoing and next generation
cosmological surveys, including, for example, the Dark Energy Survey
\citep{DES05}, the Large Synoptic Survey Telescope \citep{Ivezic08},
the \emph{Euclid} mission \citep{Laureijs11}, the
Javalambre-Physics of the Accelerated Universe Astrophysical Survey
\citep[J-PAS,][]{Benitez14} and the eROSITA mission
\citep{Merloni12}, are expected to observe many thousands of galaxy
clusters out to redshifts beyond $z\sim1$.

As such there is great demand for cluster-finding algorithms that
remain robust, reliable and efficient out to high redshift and for
catalogues of varying degrees of incompleteness. Many different
methods exist for detecting galaxy clusters in optical/near-infrared
selected surveys, as well as other approaches based on measurements of
X-ray emission \citep[e.g.][]{Ebeling00, Rosati02, Bohringer04}, weak
gravitational lensing \citep[e.g.][]{Tyson90, Bartelmann01, Leonard14}
or the Sunyaev-Zeldovich effect \citep[e.g.][]{Sunyaev72, Carlstrom02,
Ascaso07}. 

For cluster detection in optical or near-infrared
datasets, several techniques have been developed, which can be
classified broadly into three groups according to the galaxy
information that they primarily rely on. First are those approaches
based primarily upon the spatial extent of the cluster galaxies, such
as the Counts-in-Cells technique \citep[e.g.][]{Couch91, Lidman96},
Percolation algorithms \citep[e.g.][]{Huchra82, Dalton97, Eke04,
Ramella02, Robotham11} and the Voronoi-Delauney method
\citep[e.g.][]{Ramella01, Marinoni02, Kim02}, which identify clusters
as density enhancements over the mean background. The chief strength
of these algorithms is their simplicity, namely their lack of
assumptions regarding cluster shapes and their ability to work with
single-band selections. Their sensitivity to line-of-sight positions,
however, typically limits their use to spectroscopic surveys, though
there have been some attempts to apply such algorithms to photometric
datasets \citep[e.g.][]{Botzler04, Farrens11, Jian14}.

Second are the detection techniques that instead identify cluster
candidates through the presence of a red sequence; the population of
red, elliptical galaxies in clusters, typically thought to have had
their star-formation quenched by feedback processes. Assuming that the
cluster galaxy population is dominated by early-type galaxies and that
this population follows a tight colour-magnitude relation with little
intrinsic scatter, then, when imaged in two photometric bands
bracketing the $4000{\rm \AA}$ break, the cluster red sequence
galaxies will be the brightest, reddest objects
\citep{Stanford98,Gladders00}. By dividing the colour-space into
slices (according to a red sequence model) and assigning a weight to
each galaxy based upon the likelihood that the galaxy belongs to
particular slice, one can construct a surface density map for each
slice, with the peaks in the density corresponding to the cluster
candidates. Examples include the Cluster Red Sequence method
\citep{Gladders00, Lopez-Cruz04, Gladders05}, the Cut-and-Enhance
algorithm \citep{Goto02}, the MaxBCG algorithm \citep{Hansen05,
Koester07}, the C4 algorithm \citep{Miller05}, the ORCA algorithm
\citep{Murphy12} and the redMaPPer algorithm \citep{Rykoff14}. These
algorithms are popular choices for use with photometric datasets,
though there is the obvious concern that such algorithms are biased
towards those clusters with an established red sequence.

Finally, are the techniques that model characteristics of clusters,
such as the spatial or luminosity distribution of galaxies in
clusters, and test how well the galaxies in a particular region of the
sky match this model. For example, the Matched Filter technique
\citep{Postman96}, models the distribution of galaxies within a
cluster as a sum of the background density and a parametrised
function of the cluster galaxy luminosity function and the projected
radial profile of the cluster. One can then determine a likelihood for
the model parameters as a function of redshift and
luminosity. Maximising the likelihoods can therefore provide estimates
for the redshift and the total luminosity of a cluster. Several
extensions to the Matched Filter have been proposed, including the
Adaptive Matched Filter \citep{Kepner99}, the Hybrid Matched Filter
\citep{Kim02} and the three-dimensional Matched Filter
\citep{Milkeraitis10}. Recently, \citet{Ascaso12} implemented a
variation of the matched filter technique in a Bayesian
framework in order to assign to each galaxy a Bayesian probability that
there is a cluster centred on that galaxy. By additionally introducing
an optional prior for the presence of a cluster red sequence, they
were able to demonstrate the recovery of clusters with a red sequence
without the need for colour-magnitude modelling. Matched filter
methods are typically powerful techniques capable of recovering
clusters in deep, photometric redshift surveys with high completeness
and little contamination. However, their reliance on models for the
luminosity and radial profiles of clusters suggests that their results
could be model dependent and biased towards clusters displaying
similar characteristics.

In this work we describe a novel and fully Bayesian approach to detect
halos with masses above specific thresholds as peaks in the smooth
matter density field inferred from observations. To achieve this goal
we capitalise on, firstly, the previously developed \HADES{}
\citep{Jasche10b,Jasche12} large-scale structure inference framework,
which is designed to infer, from observations, the smooth
three-dimensional matter density field of the cosmic large-scale
structure, and, secondly, the Blackwell-Rao Estimator, which we use to
relate the inferred density amplitudes to the properties of dark
matter halos. Our framework exploits information from the entirety of
a galaxy survey and makes no assumptions regarding the spatial extent
of clusters, the functional form of their radial profiles or the
presence of a red sequence, which can be affected by cosmic
variance. Instead our method relies upon the more fundamental
assumption of our understanding of the matter power spectrum, which
can in turn be sampled self-consistently as part of the Bayesian
framework \citep[see e.g.][]{Jasche10a,Jasche13b,Jasche15}. To examine
the success of our methodology we make use of a realistic mock galaxy
catalogue for which halo memberships of the galaxies are known.

The layout of the paper is as follows. In $\S$\ref{sec:methodology} we
present the Bayesian inference framework \HADES{} and describe our
process for generating a realistic mock catalogue for the 6 degree
Field Galaxy Survey (6dFGS). The inference of the three-dimensional
density field for this dataset is described in
$\S$\ref{sec:application_to_mock}, followed by a discussion of
inference results. In $\S$\ref{sec:halo_detection} we describe our
approach to detect halos of different masses in observations via a
Blackwell-Rao methodology. Subsequently, we apply this approach to the
inference results obtained by the application of \HADES{} to the 6dFGS
mock catalogue and estimate its performance to recover halos in a
realistic, data driven scenario. Finally we summarise and draw
conclusions in $\S$\ref{sec:conclusions}. All magnitudes are in the
Vega system. Details of the cosmological model that we adopt are given
in $\S$\ref{sec:galform}.

\section{Methodology}
\label{sec:methodology}
In this section we first give a brief overview of the Bayesian
inference algorithm, \HADES{}, that we employ and then
introduce the N-body simulation and the semi-analytical galaxy
formation model, \GALFORM{}, that we use to construct our mock galaxy
catalogue.

\subsection{The HADES algorithm}
\label{sec:hades}

In this work we use the HAmiltonian Density Estimation and Sampling
algorithm \citep[\HADES{},][]{Jasche10b,Jasche10c,Jasche12}; a full
scale Bayesian inference framework designed to analyse modern galaxy
large-scale structure surveys on both linear and non-linear cosmic
scales, whilst simultaneously providing the corresponding uncertainty
quantification.

The three-dimensional large-scale structure of the cosmic web offers a
wealth of valuable information for testing our current picture of
cosmological structure and galaxy formation. However, connecting
observations to theoretical predictions is not trivial. Observations
of the large-scale structure are typically subject to a variety of
systematic and statistical uncertainties, such as survey geometries,
selection effects, galaxy biases, the noise of the galaxy distribution
and cosmic variance. All these effects have to be carefully accounted
for to ensure that we do not draw erroneous conclusions on the final
inferred quantities. Additional complexity for the inference of the
three-dimensional density field arises from the fact that in this work
we seek to analyse the large-scale structure on scales of
$\sim4\,\hMpc$, in the mildly non-linear and non-linear regimes. At
these scales the non-linearly evolved density field no longer obeys
simple Gaussian statistics as gravitational interactions introduce
mode coupling and phase correlations. Unfortunately there is no
tractable solution, in the form of a fully multivariate probability
distribution, for the non-linear three-dimensional density
field. There exist, however, phenomenological approximations, such as
the log-normal distribution.

The log-normal distribution can be justified via theoretical
arguments, as shown by \citet{Coles91}, and has been demonstrated to
fit, with reasonable accuracy, the one-point distributions obtained
from numerical large-scale structure simulations \citep{Kayo01}. Using
the log-normal distribution together with a suitable choice for the
cosmic power spectrum to account for one- and two-point statistics of
the density field is thus a logical choice for a prior distribution
used in Bayesian inferences of the non-linear matter
distribution. From an information theory perspective such a log-normal
prior is well justified, since it is a maximum entropy prior on a
logarithmic scale. This means that amongst all possible probability
distributions with the same mean and covariance matrix on a
logarithmic scale, the log-normal distribution is the distribution
that contains the least information. As such the log-normal
distribution represents the least informative prior for a positive
three-dimensional density field, once the mean and covariance matrix
are specified \citep[][]{Jasche10b,Jasche10c}.

To find a suitable likelihood distribution we note that the galaxy
distribution is conditionally dependent on the underlying
three-dimensional matter density field. In particular, in the most
naive picture of galaxy formation, galaxies are predominantly found in
regions of higher density than in regions of lower density. The local
noise structure of the galaxy distribution is therefore dependent on
the underlying matter density field. This feature of signal-dependent
noise is missed in traditional approaches based on Gaussian
approximations such as Wiener filtering \citep[][]{Fisher94,
Zaroubi95, Erdogdu04, Kitaura09, Jasche10a}. Assuming galaxies to be
discrete particles, their distribution can be described as a specific
realisation drawn from an inhomogeneous Poisson process, which
captures the essential features of such a signal dependent noise
\citep[see e.g.][]{Layzer56, Peebles80, Martinez02}.

Consequently, analyses of the three-dimensional density field in the
non-linear regime requires the solving of a large-scale Bayesian
inverse problem with a log-normal Poisson distribution. To explore
this highly non-Gaussian and non-linear problem the \HADES{} algorithm
relies on a Hybrid Monte-Carlo (HMC) scheme, which, instead of the
random walk behaviour displayed by traditional Metropolis-Hastings
algorithms, follows a persistent motion similar to particle
trajectories in classical mechanics problems (see \citealt{Jasche10b}
for a detailed discussion of the necessary equations of motion and
their numerical implementation). Being a fully Bayesian method, the
\HADES{} algorithm does not only provide a single estimate of the
density field but rather a full numerical representation of the
large-scale structure posterior conditional on the observations,
including a detailed treatment of all systematic and stochastic
uncertainties. The output products from \HADES{} are therefore a set
of realisations of the three-dimensional density field in a voxel
grid, as well as a measurement of the corresponding matter power
spectrum. In this fashion, the algorithm permits determination of any
desired statistical summary such as the mean, mode and variance and
simultaneously provides a straightforward means to non-linearly
propagate non-Gaussian uncertainties on any inferred quantity
\citep[][]{Jasche10b,Jasche10c}.

Recently, the \HADES{} algorithm has been extended to account for
photometric redshift uncertainties by using a block sampling procedure
\citep[][]{Jasche12}. This update means that \HADES{} is able to
account for the corresponding redshift uncertainties of millions of
galaxies observed by photometric surveys, whilst simultaneously
inferring an accurate representation of the three-dimensional density
field from such datasets. For a more detailed overview of the Bayesian
inference framework implemented in \HADES{} the interested reader is
referred to previous publications: \citet{Jasche10b, Jasche10c} and
\citet{Jasche12}.

\subsection{Generating a mock catalogue}
\label{sec:mock_catalogue}

In order to demonstrate the capability of our approach to
identify halos of galaxies, we apply \HADES{} and our halo detection
methodology to a synthetic mock galaxy catalogue in which halo
memberships are known. This will allow us to quantify how well our
approach can recover the original structures.

To this end we construct a mock catalogue to emulate the Six-degree
Field Galaxy Survey \citep[6dFGS,][]{Jones04}, which was carried out
between 2004 and 2009 using the 6-degree Field automated fibre
positioner and spectrograph system \citep[6dF,][]{Parker98,Watson00}
on the UK Schmidt Telescope at the Australian Astronomical
Observatory\footnote{Formally the Anglo Australian Observatory.}
(AAO). The 6dFGS is a near-infrared selected galaxy survey covering
the whole of the Southern sky, approximately $17,000^{\circ}$, down to
a galactic latitude of $|b|>10^{\circ}$. As of the final data release
\citep[DR3, ][]{Jones09}, the 6dFGS yielded a catalogue of
approximately 125,000 extra-galactic redshifts complete to $\left
(\band{K},\band{H},\band{J},\band{r_F},\band{b_J}\right
)=(12.65,12.95,13.75,15.60,16.75)$. Here, we construct a mock
catalogue to emulate the \band{K}-band selected sub-sample of the
6dFGS, which with approximately 93,000 redshifts constitutes the
majority of the survey.

We choose to emulate the 6dFGS for several reasons. Firstly, the 6dFGS
has a large sky coverage, with a close to uniform completeness across
the majority of the survey area. Secondly, the shallow depth of the
6dFGS means that there is little structure evolution throughout the
domain of the survey. As such, we are able to, in the first instance,
demonstrate our halo detection methodology on a density field that is
evolving very little with redshift. This means that we can approximate
the matter density field throughout the mock catalogue using the $z=0$
snapshot of the \emph{MS-W7 Simulation} \citep{Guo13}. This allows us
to provide a simple proof-of-concept of the approach. Future
application of the methodology to deeper surveys, such as the Sloan
Digital Sky Survey \citep[SDSS,][]{York00}, can then be achieved by
incorporating a more sophisticated approach to model the
redshift-dependence of the matter density field. Thirdly, we plan in
future work to apply our approach to the real 6dFGS, which contains a
rich variety of well-studied local structures, ranging from small
groups, to large super-clusters such as the Shapley Super-cluster.


\begin{figure}
  \centering
  \includegraphics[width=0.46\textwidth]{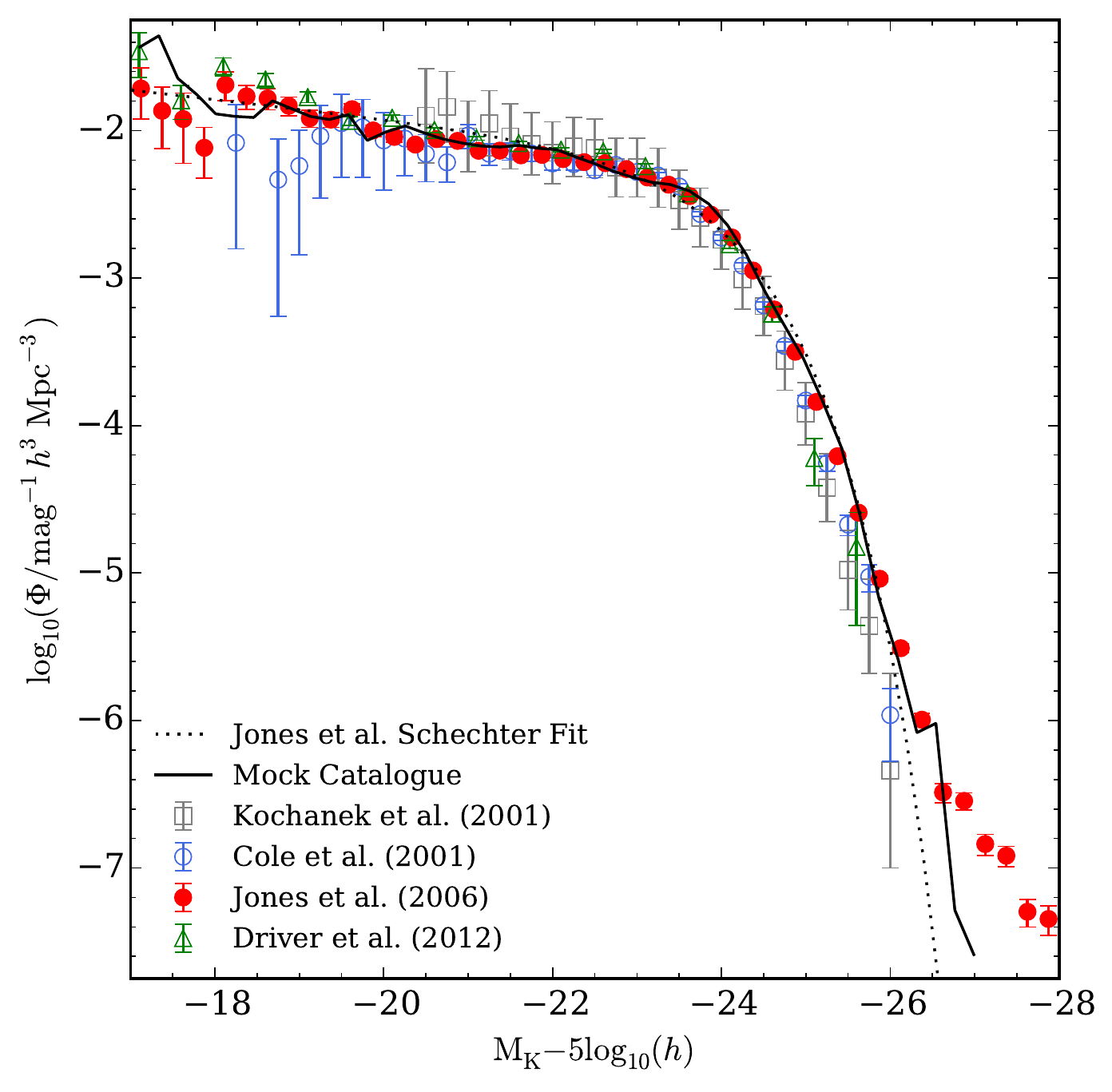}
  \caption{\band{K}-band luminosity function at $z=0$ for the
  idealised mock catalogue (solid black line). Also plotted for
  comparison are the 6dFGS \band{K}-band luminosity function estimate
  from \protect\citet{Jones06}, as well as \band{K}-band luminosity
  function estimates from \protect\citet{Kochanek01},
  \protect\citet{Cole00} and \protect\citet{Driver12}. The dotted line
  shows the \citet{Schechter76} functional fit to the 6dFGS luminosity
  function using the parameters from \protect\citet{Jones06}.}
  \label{fig:Kband_LF}
\end{figure}

\subsubsection{Galaxy formation model}
\label{sec:galform}

To construct a 6dFGS mock catalogue we follow a construction method
similar to that of \citet{Merson13}, which involves first populating
the dark matter halo merger trees of a cosmological N-body simulation
with galaxies using a semi-analytical model.

The cosmological simulation that we use is the \emph{MS-W7 Simulation}
\citep{Guo13}, which is a version of the \emph{Millennium Simulation}
\citep{Springel05} constructed using a cold dark matter (CDM)
cosmology consistent with the 7-year results of the \emph{Wilkinson
  Microwave Anisotropy Probe} \citep[WMAP7,][]{Komatsu11}. The
cosmological parameters are: a baryon matter density $\Omega_{{\rm b}}
= 0.0455$, a total matter density $\Omega_{{\rm m}} = \Omega_{{\rm b}}
+ \Omega_{{\rm CDM}} = 0.272$, a dark energy density $\Omega_{{\rm
    \Lambda}} = 0.728$, a Hubble constant $H_0 = 100h \,{\rm km\,
  s}^{-1}\Mpc^{-1}$ where $h = 0.704$, a primordial scalar spectral
index $n_{\rm s}=0.967$ and a fluctuation amplitude
$\sigma_{8}=0.810$.

The hierarchical growth of cold dark matter structure is followed at
62 fixed epoch snapshots, spaced approximately logarithmically in
expansion factor between redshift $z=127$ and the present day, in a
cubic volume of size $500\,h^{-1}\Mpc$ on a side. For each snapshot,
groups of dark matter particles are first identified through the
application of a friends-of-friends algorithm \citep{Davis85}. The
substructure-finder {\tt SUBFIND} \citep{Springel01} is then applied
to break these groups down into identifiable, self-bound
sub-halos. Independent halos are determined by establishing a sub-halo
hierarchy and identifying those sub-halos that are not bound by any
more massive sub-halos. By tracking sub-halo descendants between the
subsequent output snapshots a halo merger tree can be
constructed. Further details regarding construction of the halo merger
trees can be found in \citet{Merson13} and \citet{Jiang14}. The MS-W7
simulation uses $2160^3$ particles to represent the matter
distribution, with the requirement that a halo consists of at least 20
particles for it to be resolved. This corresponds to a halo mass
resolution of $M_{{\rm halo,lim}} \simeq
1.87\times10^{10}\,h^{-1}\Msol$, significantly smaller than expected
for the Milky Way's dark matter halo. (Within our chosen
semi-analytical galaxy formation model, halos of this mass typically
host galaxies with $M_{\band{K}}-5\log_{10}(h)\sim-11.7$).

We model the star formation and merger history of galaxies using the
\GALFORM{} semi-analytical model of galaxy formation
\citep{Cole00}. Here we adopt the recent version presented by
\citet{Gonzalez-Perez14}. The \GALFORM{} model populates dark matter
halos with galaxies using a set of coupled differential equations to
determine how, over a given time-step, the ``subgrid'' physics
regulates the size of the various baryonic components of galaxies. The
physical processes modelled by \GALFORM{} include: (i) the collapse
and merging of dark matter (DM) halos, (ii) the shock-heating and
radiative cooling of gas inside DM halos, leading to the formation of
galactic discs (iii) quiescent star formation in galactic discs, (iv)
feedback as a result of supernovae, active galactic nuclei and
photo-ionisation of the inter-galactic medium, (v) chemical enrichment
of stars and gas, (vi) dynamical friction driven mergers of galaxies
within DM halos, capable of forming spheroids and triggering starburst
events, and (vii) disk instabilities, which can also trigger starburst
events. As detailed in \citet{Merson13}, how galaxies are placed into
the dark matter halos depends on their status as central or satellite
galaxies. Central galaxies are placed at the centre of the most
massive sub-halo of their host halo. Following halo merger events,
satellite galaxies are placed at the centre of mass of what was the
most massive sub-halo of their original host halo when they were still
a central galaxy. If this sub-halo can no longer be identified, the
galaxy is placed on what was the most bound dark matter particle of
that sub-halo. The \GALFORM{} model is able to make predictions for
numerous galaxy properties, including luminosities over a substantial
wavelength range extending from the far-UV through to the
sub-millimetre.

\subsubsection{Catalogue construction}
\label{sec:mock_construction}

\begin{figure*}
  \centering
  \includegraphics[width=0.95\textwidth]{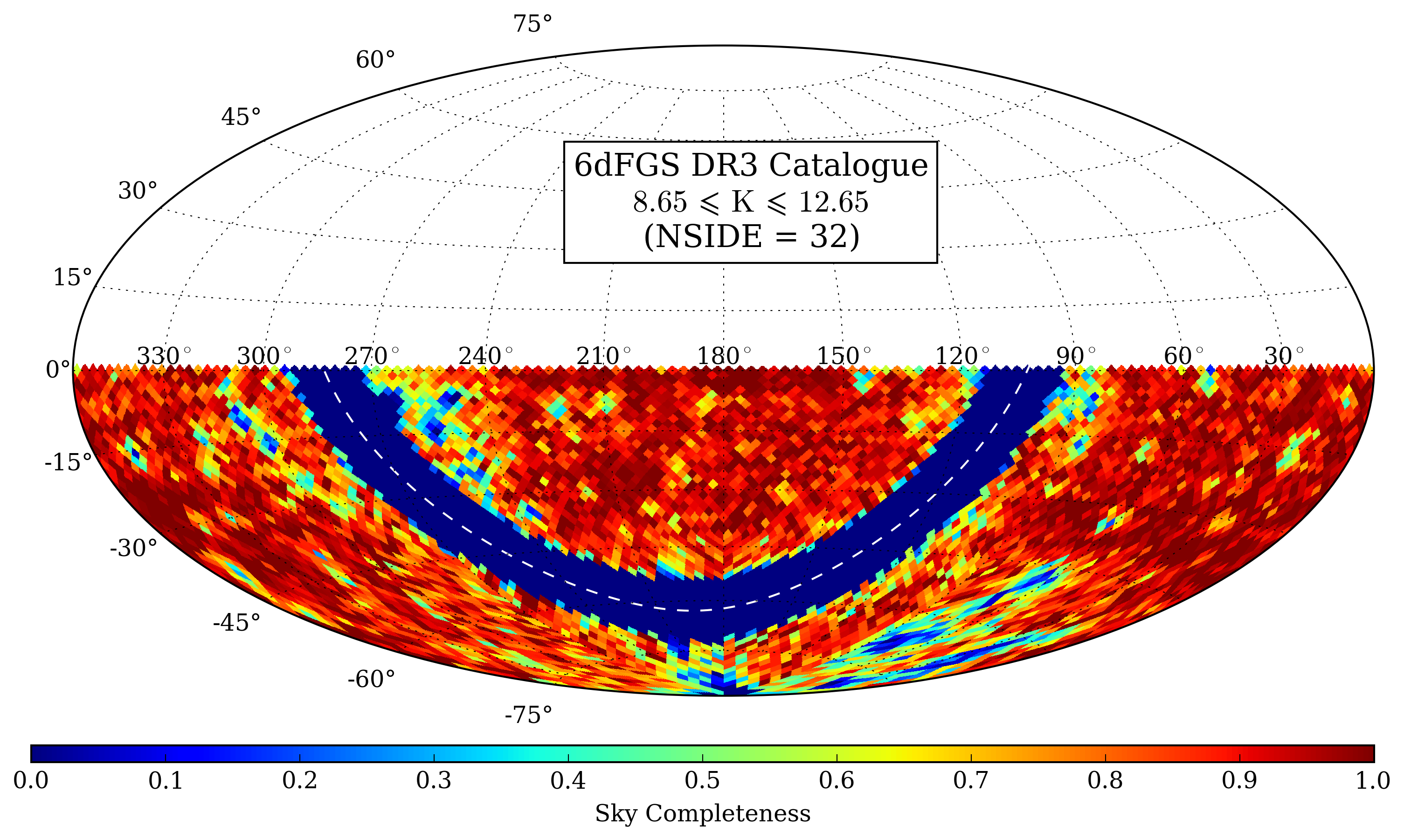}
  \caption{Redshift completeness as a function of sky position,
  $R(\mathbf{\theta})$, for the 6dFGS DR3 (in \HEALPIX{} format).}
  \label{fig:completeness_mask}
\end{figure*}

To construct the 6dFGS mock catalogue, we first run the \GALFORM{}
  model on the $z=0$ snapshot of the MS-W7 simulation. An observer is
  then placed in the box at $(\mathrm{X_o}, \mathrm{Y_o},
  \mathrm{Z_o})=(0, 0, 500)\hMpc$ and all galaxy positions are
  translated so that the observer is at the origin. To generate a
  cosmological volume comparable to that of the 6dFGS we stack a
  further three replications of the $z=0$ box such that we have a
  cuboid spanning, relative to the observer, $[-500,500]\hMpc$ in the
  X and Y directions and $[-500,0]\hMpc$ in the Z direction. Note
  that, given the cosmology of the simulation, a co-moving distance of
  $500\hMpc$ corresponds to a redshift $z\sim0.17$.

We next apply the selections to mimic the 6dFGS. Firstly we use the
  Cartesian positions of each galaxy to compute a sky position and
  redshift for that galaxy. The cosmological redshift of the galaxy is
  calculated from the co-moving distance to the galaxy from the
  observer, $r_\mathrm{com}$, defined by,
\begin{equation}
  r_{{\rm com}}(z) = {\displaystyle \int^z_0\frac{c\,{\rm
d}z^{\prime}}{H_0\sqrt{\Omega_{\rm m}\left ( 1+z^{\prime} \right )^3 +
\Omega_{{\rm \Lambda}}}}},
\label{eq:r_c(z)}
\end{equation}
where $c$ is the speed of light. For the purposes of our \HADES{}
  analysis we place an initial cut so that all galaxies with
  cosmological redshift $z>0.16$ are discarded. Note that this
  redshift is well beyond the median redshift of the 6dFGS,
  $z_{\mathrm{med}}\sim0.05$. We calculate an observed redshift,
  $z_{\mathrm{obs}}$, of each galaxy using,
\begin{equation}
z_{{\rm obs}} = \left ( 1+z \right ) \left ( 1+\frac{v_{{\rm
      r}}}{c}\right ) -1,
\label{eq:z_obs}
\end{equation}
where $v_{\mathrm{r}}$ is the radial component of the peculiar
  velocity vector, $\mathbf{\vec{v}}$, of the galaxy (i.e. $v_{{\rm
  r}}=\mathbf{\vec{v}}\cdot\mathbf{\hat{r}}$, where $\mathbf{\hat{r}}$
  is the normalised line-of-sight position vector of the galaxy). Note
  that we do not incorporate any spectroscopic redshift uncertainties
  in the mock catalogue. To mimic the solid angle footprint of the
  6dFGS we reject any galaxies with declination $\delta>0^{\circ}$ as
  well as those galaxies with a galactic latitude $|b|<10^{\circ}$.

The next step is to apply the \band{K}-band flux selection limit of
the 6dFGS, $\band{K}<12.65$, to reject those galaxies that are too
faint to have been observed. The \GALFORM{} model provides the
absolute K-band magnitude, $\mathrm{M_K}-5\log_{10}(h)$, of each
galaxy. We calculate the apparent K-band magnitude, $\mathrm{K}$, of
each galaxy using,
\begin{gather}
  \mathrm{K} = \mathrm{M_K} -5\log_{10}(h)+ 5\log_{10}\left ( \frac{d_L\left (
z\right )}{10{\rm pc}} \right )\nonumber\\- 2.5\log_{10}\left ( 1+z\right ) +
k(z),
\label{eq:bcdm}
\end{gather}
where $d_L$ is the luminosity distance to the galaxy and $k(z)$ is an
applied K-band k-correction, which we obtain by interpolating the
tabulated k-corrections from \citet{Poggianti97}. In
Fig.~\ref{fig:Kband_LF} we show the K-band luminosity function for the
mock catalogue, which we compare with the 6dFGS K-band luminosity
function estimated by \citet{Jones06}. Note that \citeauthor{Jones06}
corrected their estimate of the 6dFGS luminosity function for
incompleteness. Our mock catalogue gives a galaxy number density that
is in excellent agreement with that of the 6dFGS, particularly around
the characteristic magnitude,
$\mathrm{M_K^{\ast}}-5\log_{10}(h)=-23.83$.

At this stage, the mock catalogue that we have represents an idealised
copy of the 6dFGS, such that the catalogue is complete down to the
flux limit and complete over the extent of the 6dFGS DR3 footprint on
the sky. The final step is to degrade the completeness of our
idealised mock catalogue such that we model the effect of systematics
that are introduced into observational datasets as a result of survey
strategy. For spectroscopic surveys such as the 6dFGS, incompleteness
is introduced as a result of observational limitations, such as fibre
collisions and effects of poor observing conditions, which prevent one
from obtaining a redshift measurement for each target. Collisions of
the 6dF fibres, for example, prevent simultaneous observation of
galaxies with a proximity less than approximately $5.71$ arcminutes on
the sky \citep{Campbell04}, though this can be mitigated somewhat by
repeat observations. Such systematics can therefore lead to the
observed galaxy counts in any particular dark matter halo being
incomplete, which reduces the signal-to-noise of that halo. Therefore
it is important to ensure that we are applying our methodology to a
mock dataset that is representative of observational datasets and
their inherent systematics.

\citet{Jones06} model the total completeness,
$T(\boldsymbol\theta,m)$, for each galaxy in the 6dFGS using the
separable function $T(\boldsymbol\theta,m)=S(\boldsymbol\theta)C(m)$,
where $C(m)$ is the completeness as a function of magnitude, $m$, and
$S(\boldsymbol\theta)$ is a constant scaling the completeness of the
field in which a galaxy was observed to the completeness,
$R(\boldsymbol\theta)$, on that part of the sky. To remove incomplete
regions from their final dataset, \citeauthor{Jones06} selected those
galaxies for which $T(\boldsymbol\theta,m)\geqslant0.6$. In order to
fully emulate the 6dFGS we would need to mimic the observational
design of the survey, including optimally tiling the mock catalogue
with a set of 6-degree fields and modelling effects such as fibre
collision. However, given that the purpose of our mock catalogue is to
help provide a simple demonstration of the ability of the our halo
detection methodology and that to do this the mock catalogue does not
need to be a perfect emulation, we choose to adopt a simpler, more
straightforward implementation. We therefore degrade the mock
catalogue using a \HEALPIX{} \citep{Gorski05} realisation of the sky
completeness mask of the DR3 dataset, as shown in
Fig.~\ref{fig:completeness_mask}, where the colour-bar indicates the
value for the sky completeness $R(\boldsymbol\theta)$, at the sky
position, $\boldsymbol\theta$, of each \HEALPIX{} pixel. To degrade
the mock catalogue we simply use random number generation to accept or
reject galaxies based upon the value of $R(\boldsymbol\theta)$ for the
pixel to which the galaxy is assigned. By degrading the catalogue in
this way, we ensure that the sky completeness mask in
Fig.~\ref{fig:completeness_mask} is a good description for the
completeness of the mock sky. Following this procedure, we are left
with a mock catalogue that provides a reasonable approximation for a
\band{K}-band selected 6dFGS-like galaxy survey. Note that our
approach does not introduce any magnitude incompleteness,
i.e. $C(m)=1$, and instead would lead to
$T(\boldsymbol\theta,m)=T(\boldsymbol\theta)=R(\boldsymbol\theta)$.

\begin{figure}
  \centering
  \includegraphics[width=0.46\textwidth]{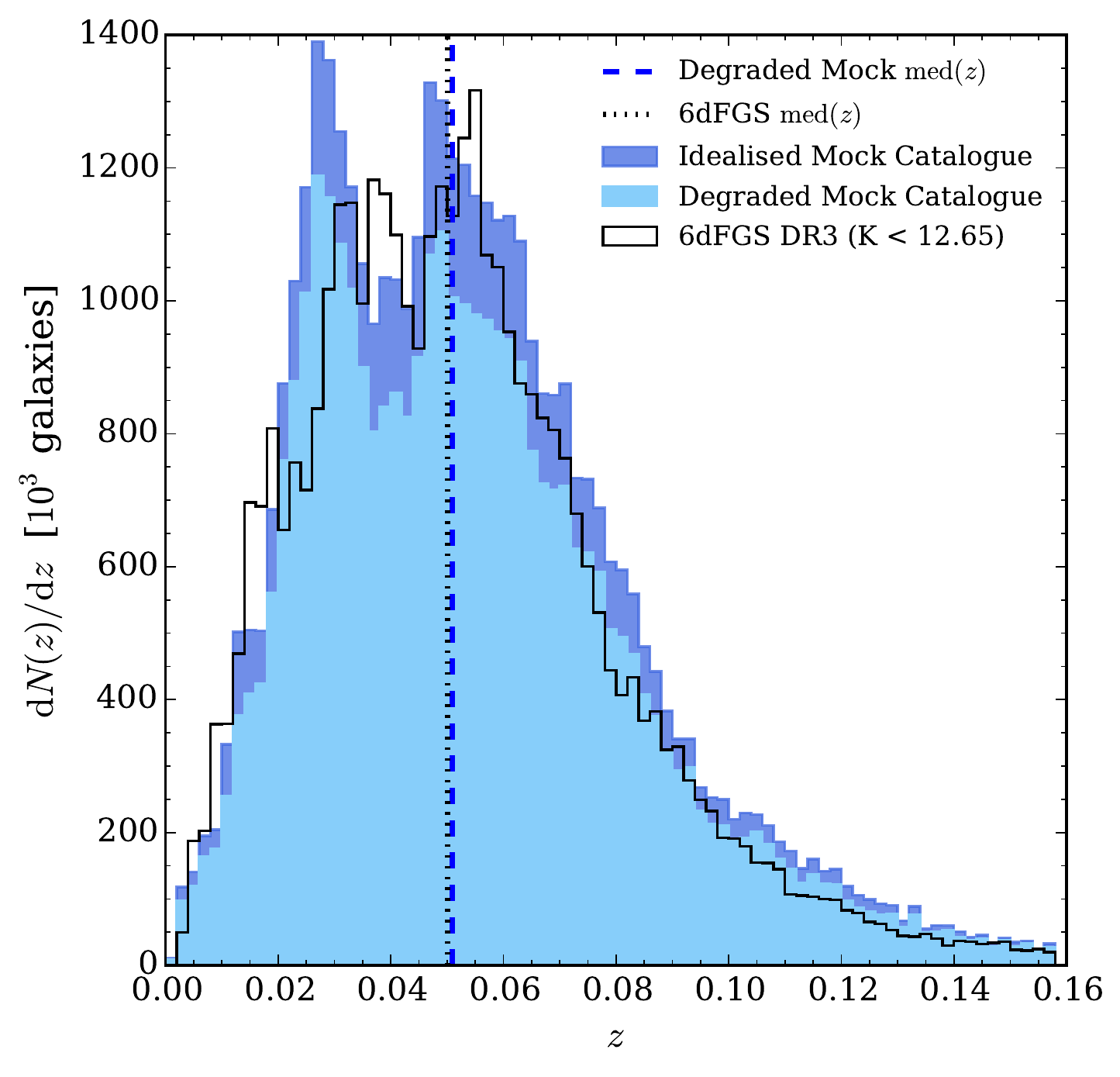}
  \caption{Redshift distributions for the idealised mock catalogue
  (dark blue shaded histogram) and the completeness degraded mock
  catalogue (light blue shaded histogram). Shown for comparison is the
  distribution for the 6dFGS DR3 \band{K}-band selected sample (black
  line). The dotted line indicates the median redshift for the 6dFGS
  DR3 galaxies, whilst the dashed line shows the median redshift for
  the degraded mock catalogue.}
  \label{fig:redshift_distribution}
\end{figure}

After degrading the mock catalogue, we are left with approximately
70,000 galaxies with a median redshift of approximately 0.05, which is
consistent with the median redshift of the 6dFGS DR3. The redshift
distributions of both the idealised and the degraded mock catalogue
are shown in Fig.~\ref{fig:redshift_distribution}. For comparison the
redshift distribution of the K-band selected 6dFGS DR3 dataset, which
constitutes about 75,000 galaxies, is also shown.


\section{Inference of the cosmic large-scale structure}
\label{sec:application_to_mock}
In this section we describe the set-up and inference results of
  applying the \HADES{} algorithm to our 6dFGS mock catalogue.


\subsection{Application of the HADES algorithm}
\label{sec:Bayesian_analysis}
As stated previously, in this work we rely on the Bayesian inference
algorithm \HADES{} to recover the three dimensional large-scale
structure from the mock observations. In particular we follow a
procedure similar to that described in \citet{Jasche10b} and
\citet{Jasche13a}.

\begin{figure}
  \centering
  \includegraphics[width=0.47\textwidth]{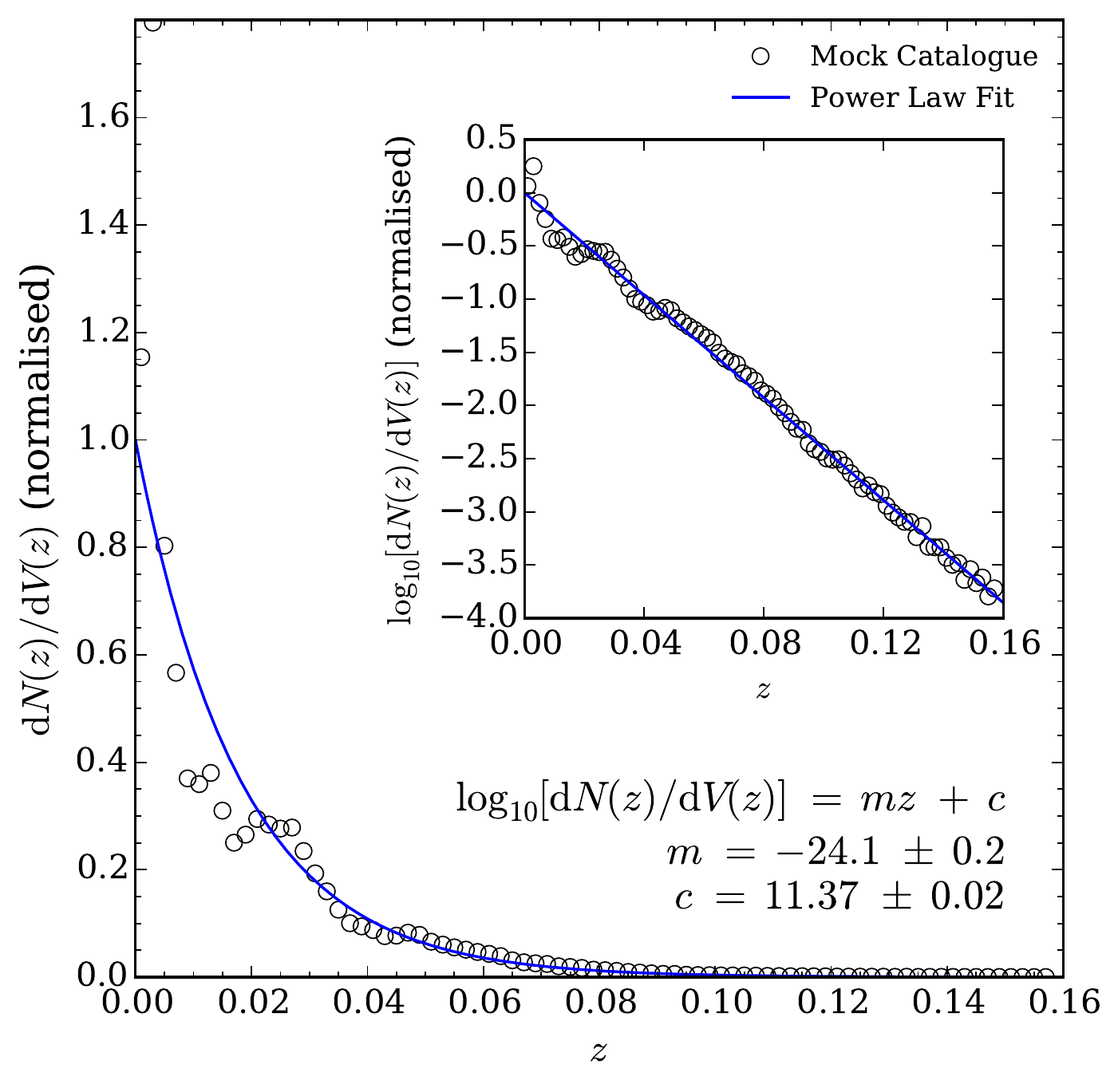}
  \caption{The volume-weighted redshift distribution the 6dFGS mock
  catalogue (open circles). The solid line shows the power fit to this
  distribution, which is provided to \HADES{} as the estimate for the
  radial selection function of the mock catalogue. The inset panel
  shows the base-10 logarithm of the distribution. Stated in the plot
  are the values for the parameters, $m$ and $c$, for the power law
  fit.}
  \label{fig:selection_function}
\end{figure}

As inputs, \HADES{} requires only the galaxy positions, the sky
completeness mask (in \HEALPIX{} format) and an estimate of the radial
selection function of the mock catalogue. We calculate the radial
selection function of the mock catalogue by computing the volume
weighted redshift distribution, ${\rm d}N(z)/{\rm d}V(z)$, which we
show in Fig.~\ref{fig:selection_function}. Remarkably this function is
very well described by a power-law, which is also shown. This
power-law relation, re-normalised to the interval [0,1], is the
selection function provided to \HADES{}. \HADES{} uses a convolution
of the sky completeness mask and the radial selection function to
construct a three-dimensional response operator,
$R(\overrightarrow{x})$, which describes the completeness of the
observations as a function of position, $\overrightarrow{x}$. For
details on the data model and the implementation of the \HADES{}
algorithm we refer the interested reader to
\citet{Jasche10b,Jasche10c} and \citet{Jasche12}.

\begin{figure*}
  \centering
  \includegraphics[width=0.92\textwidth]{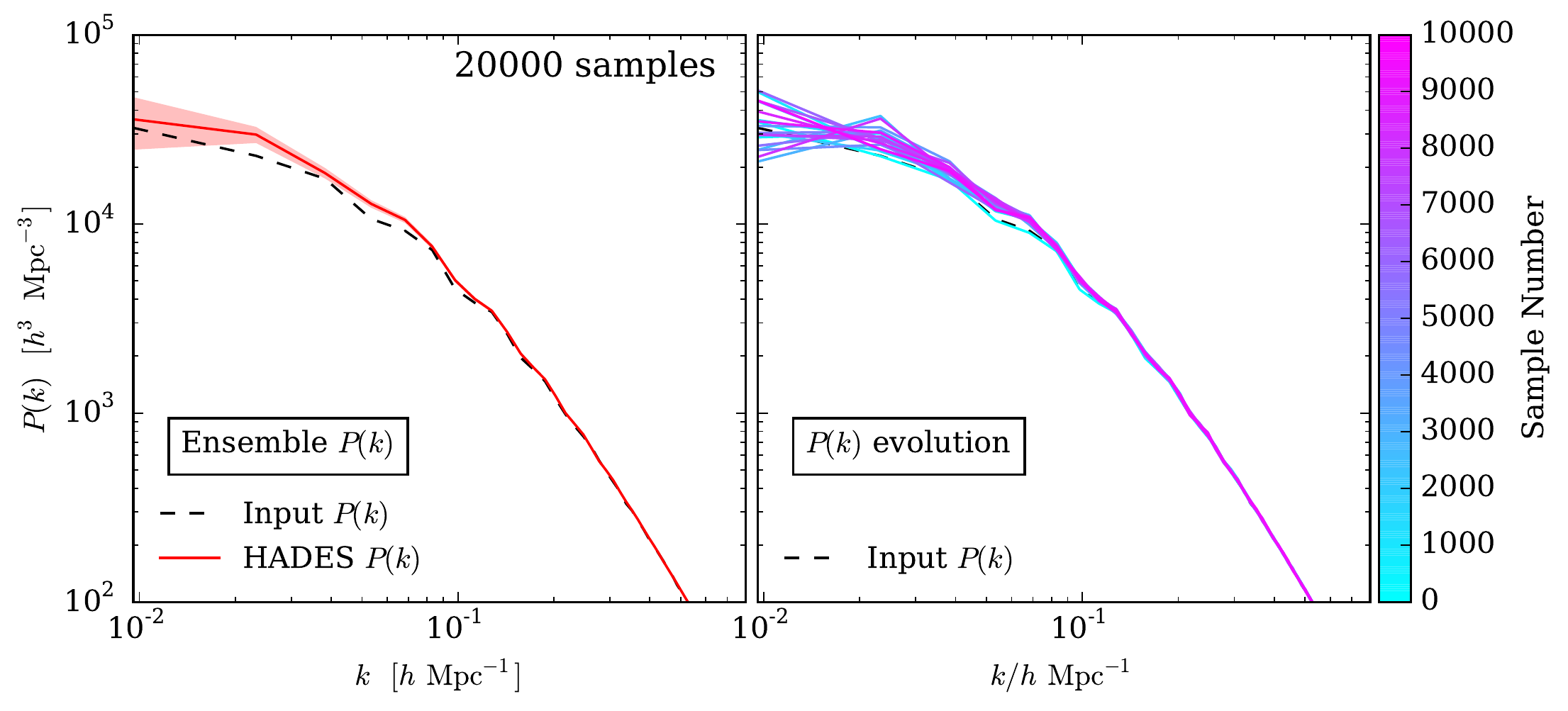}
  \caption{The matter power spectrum as recovered by \HADES{}. The
  left-hand panel shows the ensemble mean power spectrum, obtained by
  averaging over 20,000 samples, with the shaded regions indicating
  the size of the standard deviation in each bin of wavenumber. The
  right-hand panel shows the evolution of the power spectrum with
  sample number for one \HADES{} chain. The solid line for each
  estimate is coloured according to the number of the sample it was
  taken from. In each panel, the dashed line corresponds to the input
  power spectrum that \HADES{} was provided with.}
  \label{fig:mock_pk}
\end{figure*}

We infer the large-scale structure within a rectangular Cartesian
domain of size length $981\hMpc \,\times\, 955\hMpc \,\times\,
511\hMpc$. This inference domain was chosen to optimally account for
the geometry of the 6dFGS mock catalogue. The inference domain was
subdivided into $256 \,\times\, 256 \,\times\, 128$ cells, allowing a
grid resolution of $\sim 3.6\hMpc$. We note that the total number of
inference parameters, which correspond to the density amplitudes in
each of the grid cells, is $\sim 10^6$. This large number of
parameters can be efficiently sampled by the \HADES{} algorithm via a
Hamiltonian Monte Carlo sampling framework. To explore the
corresponding high dimensional parameter space we run four chains in
parallel, each generating a total of $10,000$ data constrained
realisations of the three-dimensional density field. Being a numerical
representation of the full posterior distribution, this ensemble of
density fields contains all of the information that could be extracted
from observations and provides accurate quantification of
uncertainties inherent to any cosmological observation.

\subsection{Burn-in and statistical efficiency}
\label{sec:burn_in}

As with any Markov Chain Monte-Carlo technique, there will be
correlations between subsequent density field realisations generated
by the Markov chain. For this reason the sampler requires a certain
amount of sampling steps to decorrelate from the chosen initial
conditions. This phase of a Markov sampler is referred to as the
burn-in period. After this finite initial phase the Markov sampler
generates density field realisations drawn from the correct target
posterior distribution.

A simple monitor of burn-in is to follow the evolution of parameters
with sample number \citep[e.g.][]{Eriksen04,Jasche10a}. The right-hand
panel of Fig.~\ref{fig:mock_pk} shows the evolution of the recovered
posterior matter power spectrum with sample number for the 10,000
samples in one of the four Markov chains. We can see that the chain
has converged after approximately 2000 samples and starts exploring
the parameters within the range of uncertainty. As a conservative
measure, we discard the first 5000 samples in each chain to ensure
that each chain has passed the initial burn-in phase. This leaves us
with 5000 realisations of the density field for each chain, giving a
total of 20,000 samples.

The left-hand panel of Fig.~\ref{fig:mock_pk} shows the ensemble mean
and variance on the power spectrum, obtained by averaging over the
20,000 converged samples. At small $k$ the power spectrum is biased
high relative to the input power spectrum, likely due to the effect of
galaxy bias. In its current form \HADES{} assumes a constant linear
bias. We assume an arbitrary bias value of 1.2, which, given the value
of $\sigma_8$ used in the MS-W7 cosmology, is within $2\sigma$ of the
bias estimates of \citet{Beutler12}. Another possible source of the
excess power could be the appearance of repeated structures in the
mock catalogue, arising due to our method of building the mock
catalogue by replicating the simulation box.

\subsection{Inferred density fields}
\label{sec:density_field}

\begin{figure*}
  \centering
  \includegraphics[width=0.98\textwidth]{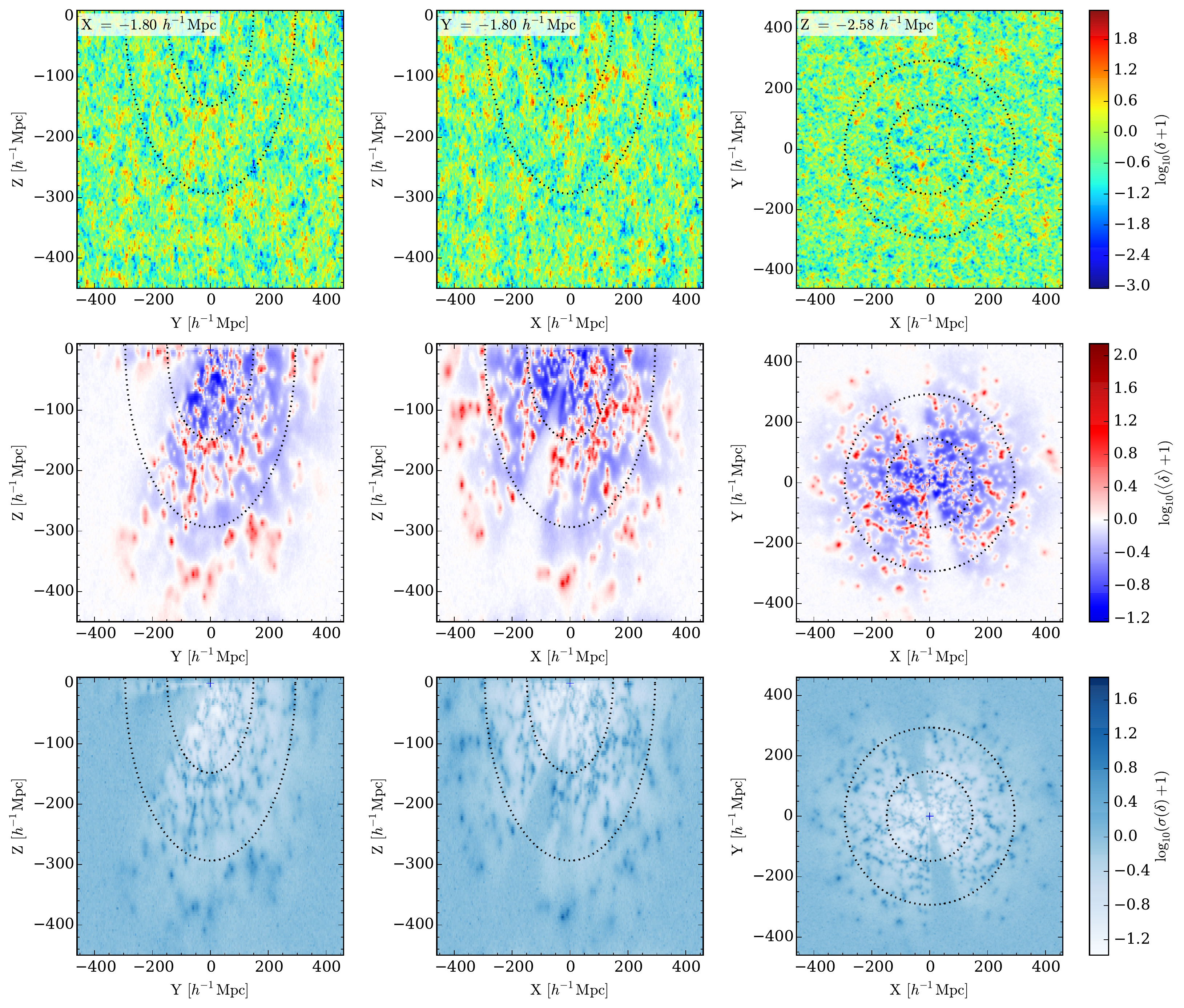}
  \caption{Slices showing the \HADES{} density field in the three
    Cartesian axes. The resolution of the \HADES{} reconstruction is
    approximately $4\,\hMpc$. The left-hand column shows a slice at
    $X\sim0\,\hMpc$, the middle column shows a slice at
    $Y\sim0\,\hMpc$ and the right-hand at $Z\sim-3\,\hMpc$. The top
    row shows a single realisation of the \HADES{} density field. The
    middle row shows the ensemble average of the density field,
    obtained by averaging over 20,000 realisations. The bottom row
    shows the ensemble variance, again obtained by averaging over
    20,000 realisations.}
  \label{fig:mock_density_field}
\end{figure*}

We now examine the density field as inferred by \HADES{}. In
Fig.~\ref{fig:mock_density_field} we show slices, of approximately
$4\,\hMpc$ thickness, through the \HADES{} density field. The
different columns correspond to a slice through each of the Cartesian
axes. In the X and Y axes the slices are approximately at the origin,
whilst the slice along the Z axis corresponds approximately to ${\rm
Z}\sim-3\,\hMpc$. (This corresponds to the slice along the Z axis that
is closest to the observer and whose volume is entirely spanned by the
mock galaxy data). The top row shows slices through a single
realisation of the recovered density field, whilst the middle row
shows the same slices through the ensemble mean density field,
$\langle \delta \rangle$, averaged over 20,000 samples. In the bottom
row we show the ensemble variance of the recovered density field,
$\sigma(\delta)$, again taken over 20,000 samples. From
Fig.~\ref{fig:mock_density_field} we can see that for many regions in
the inferred large-scale structure the ensemble variance is comparable
to the ensemble mean, as expected for a Poisson process.

Comparing these results we can see that whilst the density field from
individual samples appears very Gaussian, the ensemble mean density
field is strikingly non-Gaussian, with the non-linear features of the
cosmic web becoming clearly visible above the noise. High
signal-to-noise structures, such as galaxy clusters and voids, are
easily identifiable out to distances of approximately $200\,\hMpc$ from
the observer, which for our cosmological model corresponds to a
redshift of $z\sim0.07$. 

Note, however, that the masked regions, which are not constrained by
observations and regions dominated by noise tend towards the mean
density with $\langle\delta\rangle=0$. This behaviour is expected in
regions without data constraints, where we expect to recover the
cosmic mean density on average. One such example is the region of the
mock survey masked by the galactic plane, which is not visible in
individual realisations but becomes apparent in the ensemble
properties. In each individual sample \HADES{} is able to infer the
large-scale structure in these regions, however the lack of
constraints for these regions leads to a low signal-to-noise ratio for
the inference in these regions so that over the ensemble 20,000
realisations the inferred density field averages out to the mean
density.

\subsection{Recovery of structures}
\label{sec:recovery_of_structures}

\begin{figure*}
  \centering
  \includegraphics[width=0.95\textwidth]{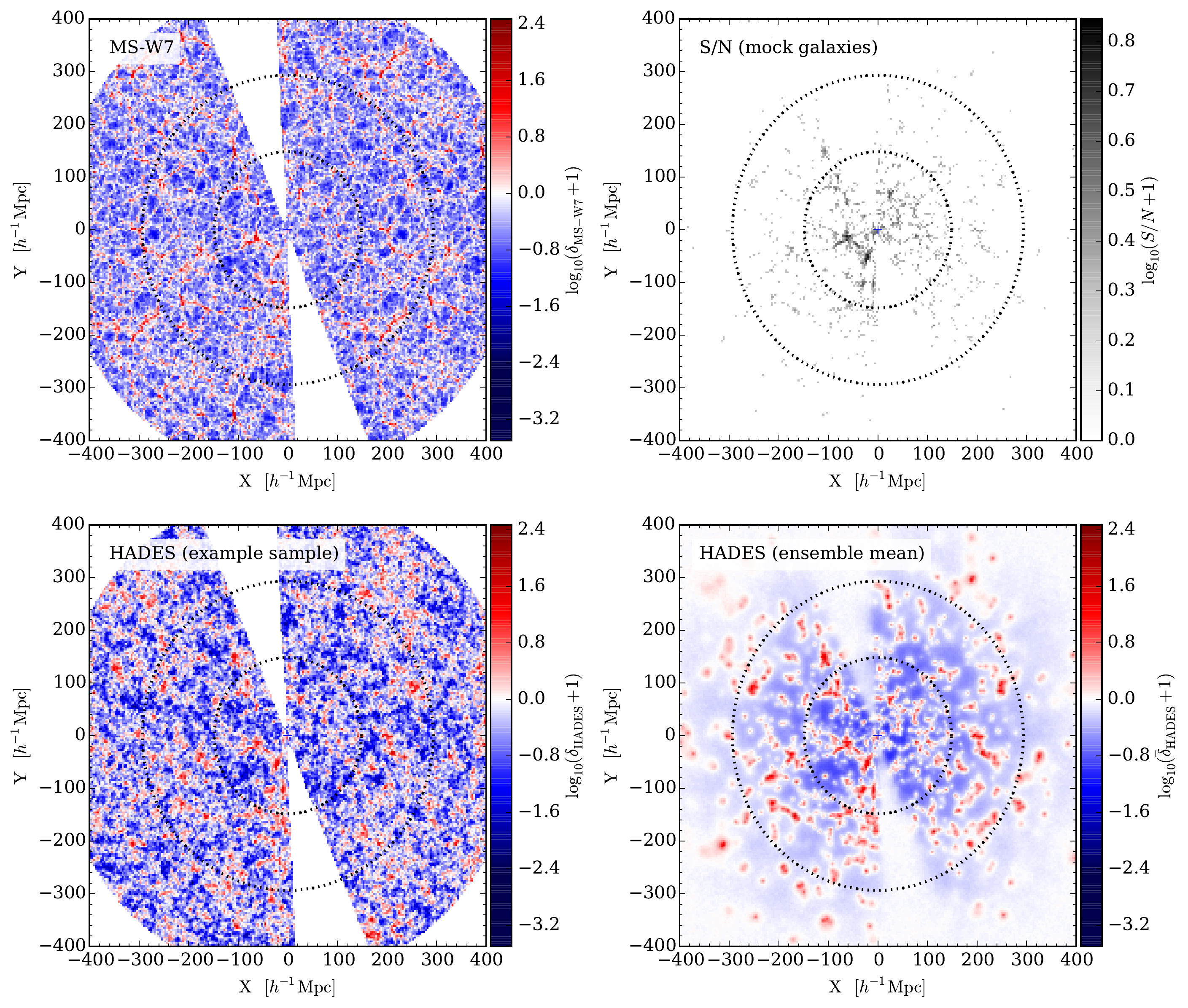}
  \caption{Zoomed slices through the \HADES{} volume at ${\rm
  Z}\sim-3\,\hMpc$ showing the MS-W7 density field in the original
  mock catalogue (top left), the signal-to-noise (S/N) ratio of the
  mock observations (corresponding to the square root of the counts,
  top right), an example \HADES{} realisation of the matter density
  field (bottom left) and the ensemble mean density field from
  \HADES{} (bottom right). Note that in the left-hand panels, the
  density field is only shown for voxels where the response operator,
  $R$, is non-zero (i.e. where the completeness of the observations is
  non-zero). The dotted concentric circles correspond approximately to
  the median redshift and twice the median redshift of the mock
  galaxies.}
  \label{fig:density_fields}
\end{figure*}

Having seen that \HADES{} is able to provide a realistic realisation
for the cosmic web, we now consider the recovery of individual
structures. We stress that the density field inferred by \HADES{}
corresponds to the continuous matter density field and that \HADES{}
does not provide any information for individual, discrete structures
or for the halo density field. It does, however, provide insight into
which individual structures, in particular clusters, could be
identified as peaks in the inferred ensemble mean density field.

In Fig.~\ref{fig:density_fields} we compare an example density field
realisation from \HADES{}, as well as the ensemble density field, with
the true density field for the mock catalogue, which corresponds to
the density field from the $z=0$ snapshot of the MS-W7 simulation. We
estimate the MS-W7 density field by replicating the MS-W7 box such
that we can count the number of dark matter particles in each of the
voxels in the \HADES{} volume. Note that for the MS-W7 density field
and the example \HADES{} realisation, we only show the density field
for voxels where the response operator, $R$, is non-zero (i.e. for
voxels where the completeness of the observations is non-zero). Hence
very distant regions, as well as regions behind the Galactic plane,
are masked out. In addition, we also show in
Fig.~\ref{fig:density_fields} the signal-to-noise ratio (S/N) for the
observations, which we estimate as the square root of the galaxy
counts in each \HADES{} voxel.

A visual comparison of the MS-W7 density field with the
\HADES{} density fields, either the example realisation or the
ensemble mean, shows that \HADES{} is recovering the large-scale
structure of the MS-W7 density field quite well, particularly for
structures within twice the median redshift of the mock galaxies (as
indicated by the outer of the two concentric circles). Individual
structures in the MS-W7 density field can be identified in the
\HADES{} density fields. For example, the structure located near the
observer at $(\mathrm{X},\mathrm{Y})\sim(-50,-20)\hMpc$, which is
clearly visible in the galaxy counts, can be readily identified in
both the \HADES{} example realisation and the ensemble mean density
field. Other structures further away from the observer, such as the
filamentary structures at
$(\mathrm{X},\mathrm{Y})\sim(-160,-140)\hMpc$ or
$(\mathrm{X},\mathrm{Y})\sim(200,-40)\hMpc$, are not easily visible in
the galaxy counts but are recovered by \HADES{}, albeit at poorer
resolution. At distances around twice the median redshift, or beyond,
only a few individual clusters can be resolved, thanks to the counts
of bright cluster galaxies. It is indeed noticeable that the fine
filamentary structure in the MS-W7 density field is less well resolved
by \HADES{} compared to galaxy clusters, which constitute the nodes of
the cosmic web. This, for example, could well be due to the fact that
\HADES{} is having to infer the density field using galaxies in
redshift-space, which will lead to individual structures being smeared
out by redshift-space distortion effects.

We note that for our analysis with \HADES{} we have neglected the
impact of uncertainties on the spectroscopic galaxy redshifts, which
are not modelled in our mock catalogue. If we examine the redshift
uncertainties, $\delta z$, of galaxies in the \band{K}-band selected
sub-sample of the 6dFGS DR3, we find that the median fractional
uncertainty is $\delta z/z=0.003^{+0.004}_{-0.001}$. (Uncertainties on
the median value correspond to the difference between the median and
the $\mathrm{10^{th}}$ and $\mathrm{90^{th}}$ percentiles). Assuming
our given cosmology, we can convert this to a fractional uncertainty
on the co-moving distance, $r$, of the galaxies, $\delta r/r$, where
we take $\delta r=[r(z+\delta z)-r(z-\delta z)]/2$. This yields a
typical fractional uncertainty of $\delta
r/r=0.003^{+0.005}_{-0.001}$. For a galaxy at the median redshift of
our mock survey, $z_{{\rm med}}\sim0.05$, this corresponds to a
typical uncertainty on the co-moving distance of approximately
$\sim0.45\,\hMpc$, which we note is much smaller than our grid
resolution of $\sim3.6\,\hMpc$ and so should have negligible impact on
our results. If we apply our methodology to a catalogue of photometric
redshifts, however, the impact from photometric redshift uncertainties
would need to be considered.

\begin{figure}
 \centering
 \includegraphics[width=0.47\textwidth]{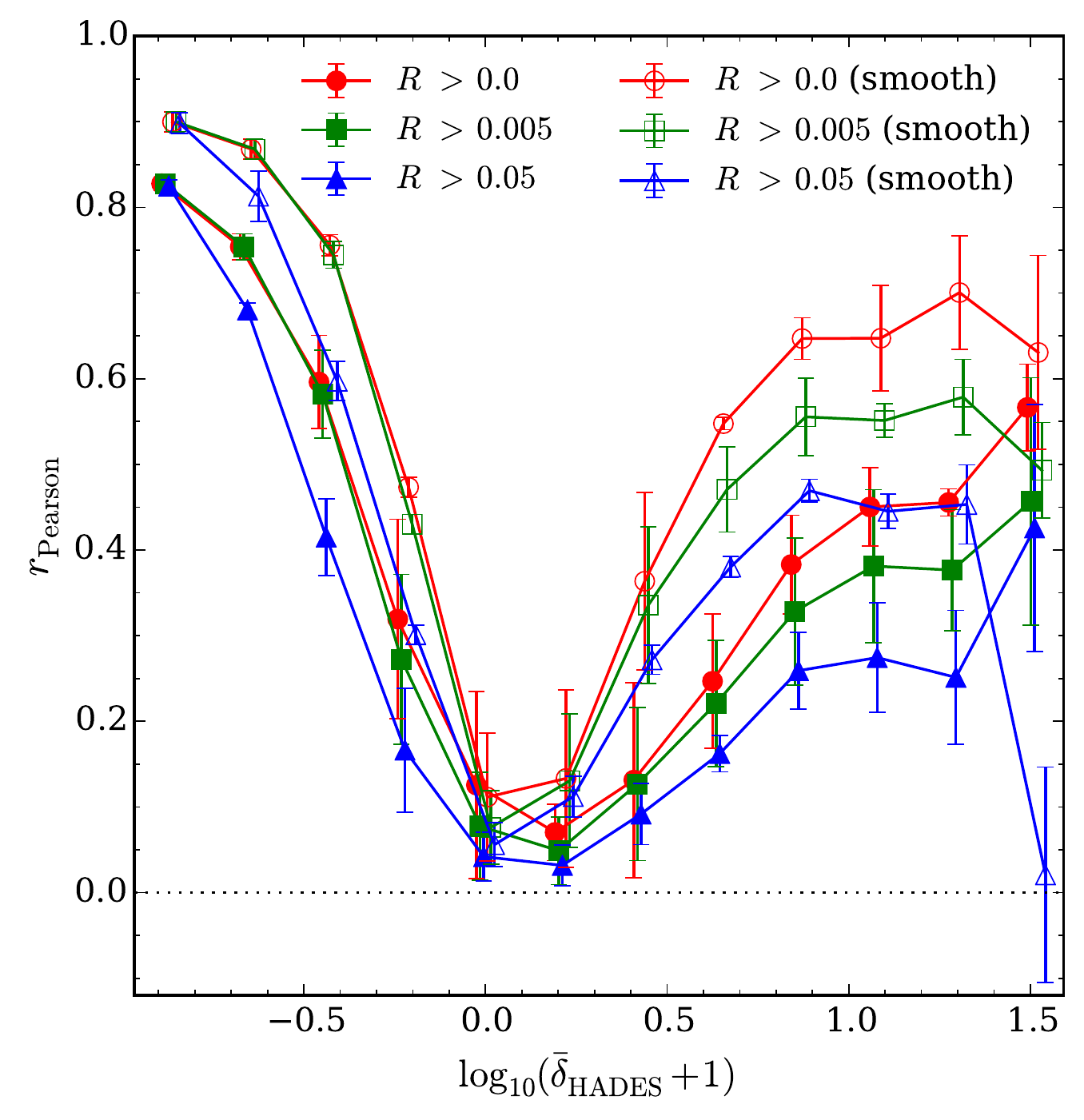}
 \caption{Pearson rank correlation coefficient indicating the mean
 strength of the correlation between the density field of the MS-W7
 simulation and each of 20,000 \HADES{} density field
 realisations. The correlation coefficient is shown as a function of
 density contrast from the ensemble mean of the \HADES{} recovered
 density fields. The points show the mean coefficient for each density
 bin and the errorbars show one standard deviation. The filled symbols
 show the correlation obtained when considering only voxels for which
 the response operator, $R$, is greater than a threshold value: 0.0
 (red circles), 0.005 (green squares) and 0.05 (blue triangles). The
 empty symbols show the correlation obtained when the density fields
 are first smoothed on scales of $\sim18\hMpc$. (The resolution in the
 non-smoothed case is $\sim3.6\hMpc$.)}
\label{fig:hades_halos_corrcoeff}
\end{figure}

To quantify our ability to recover of individual structures with
\HADES{}, we examine the correlation between the MS-W7 density field
and the density field of the \HADES{} realisations. To do this, we
measure the Pearson correlation coefficient, which varies between
$\pm1$ and provides a measure of the linear correlation between two
quantities, with $+1$ indicating a perfect positive correlation, $-1$
indicating a perfect negative correlation and $0$ indicating no
correlation. We can therefore use the Pearson correlation coefficient
to search for correlation between the true and inferred density
fields. As such, we estimate the correlation between the MS-W7 density
field and each of the individual 20,000 \HADES{} realisations,
i.e. giving us 20,000 estimates for the correlation. However, in each
case instead of obtaining a single value for coefficient over the
entire set of voxels, we split the voxels into density bins according
to the density amplitude that that voxel has in the \HADES{} ensemble
mean density field. When measuring the coefficients we only consider
voxels in the \HADES{} volume where the response operator, $R$, is
non-zero (as in the left-hand panels of
Fig.~\ref{fig:density_fields}).

In Fig.~\ref{fig:hades_halos_corrcoeff} we show the correlation
coefficient as a function of the ensemble mean density from
\HADES{}. The filled circles show the mean correlation coefficient in
each bin and the errorbars indicate one standard deviation. As can be
seen, the correlation coefficient increases to larger positive values
in the lowest and highest density bins, indicating that \HADES{} is
correctly identifying the most over-dense and under-dense voxels in
the \HADES{} grid, which correspond to the regions of highest
signal-to-noise. The correlation is higher for the lowest density
bins, which correspond to voids, than for the highest density bins,
which correspond to clusters. This is likely due to voids having a
larger volume filling factor than clusters and so being more easily
identified in lower resolution reconstructions. In addition, due to
their larger volume, the positions of the void centres will be less
affected by redshift-space distortions than to the positions of
clusters. As a consequence, large-scale structure inference algorithms
have previously been used to identify and examine the properties of
cosmic voids \citep[e.g.][]{Leclercq15, Lavaux16}.  Towards the mean
density, $\langle\delta\rangle\sim0$, the correlation weakens
significantly. This is understandable given that this density contrast
will be associated with the regions of lowest signal-to-noise, such as
the halos of small galaxy groups or even individual galaxies, where
\HADES{} is unable to make a decisive statement.

We show the correlation for two additional thresholds in
the response operator: $R>0.005$ and $R>0.05$. These increasing limits
of $R$ essentially limit us to smaller and smaller volumes about the
observer: $R>0.005$ limits us to a spherical volume within
approximately twice the median redshift and $R>0.05$ limits us to a
spherical volume within approximately the median redshift (excluding,
in all instances, the region behind the Galactic plane). Considering
the highest density bins, the correlation decreases as the limit in
$R$ is increased. Also, the uncertainty on the correlation also
increases as we are restricted to a smaller volume. These results are
consistent with the increasing impact of small-scale redshift-space
distortions, which are more prominent closer to the observer and would
shift the apparent positions of clusters in the \HADES{}
reconstructions, thus leading to a reduction in the correlation.
Furthermore, we would expect the uncertainty to increase as we
consider smaller volumes with a lower number statistics of clusters.

As a final demonstration, we also examine the impact on the
correlation of smoothing the \HADES{} and MS-W7 density fields. In
Fig.~\ref{fig:hades_halos_corrcoeff} the empty points show the
correlation coefficients obtained when the the \HADES{} and MS-W7
density fields are first smoothed using a 3-dimensional Gaussian
kernel\footnote{We adopt the Gaussian filter from the Python {\tt
Scipy} library, \url{scipy.org/}.}, adopting a $5\times5\times5$ pixel
window function. Given the pixel resolution, this window function has
a scale of approximately $18\hMpc$. As such, this smoothing will
remove all small-scale resolution but will allow us to consider
whether the \HADES{} and MS-W7 density fields correlate on
large-scales. We see in Fig.~\ref{fig:hades_halos_corrcoeff} that
smoothing the density fields in this way leads to an increase in the
correlation in the majority of the highest density bins for each of
the $R$ limits considered. Thus, we can conclude that the \HADES{}
density fields correlate well with density field from the MS-W7 on
both small-scales ($\sim4\hMpc$) and large-scales
($\sim18\hMpc$). This result strongly supports the use of \HADES{}
density field realisations for identification of galaxy clusters (and
voids) in galaxy survey datasets.


\section{Bayesian Halo Detection}
\label{sec:halo_detection}

\begin{figure*}
  \centering
  \includegraphics[width=0.95\textwidth]{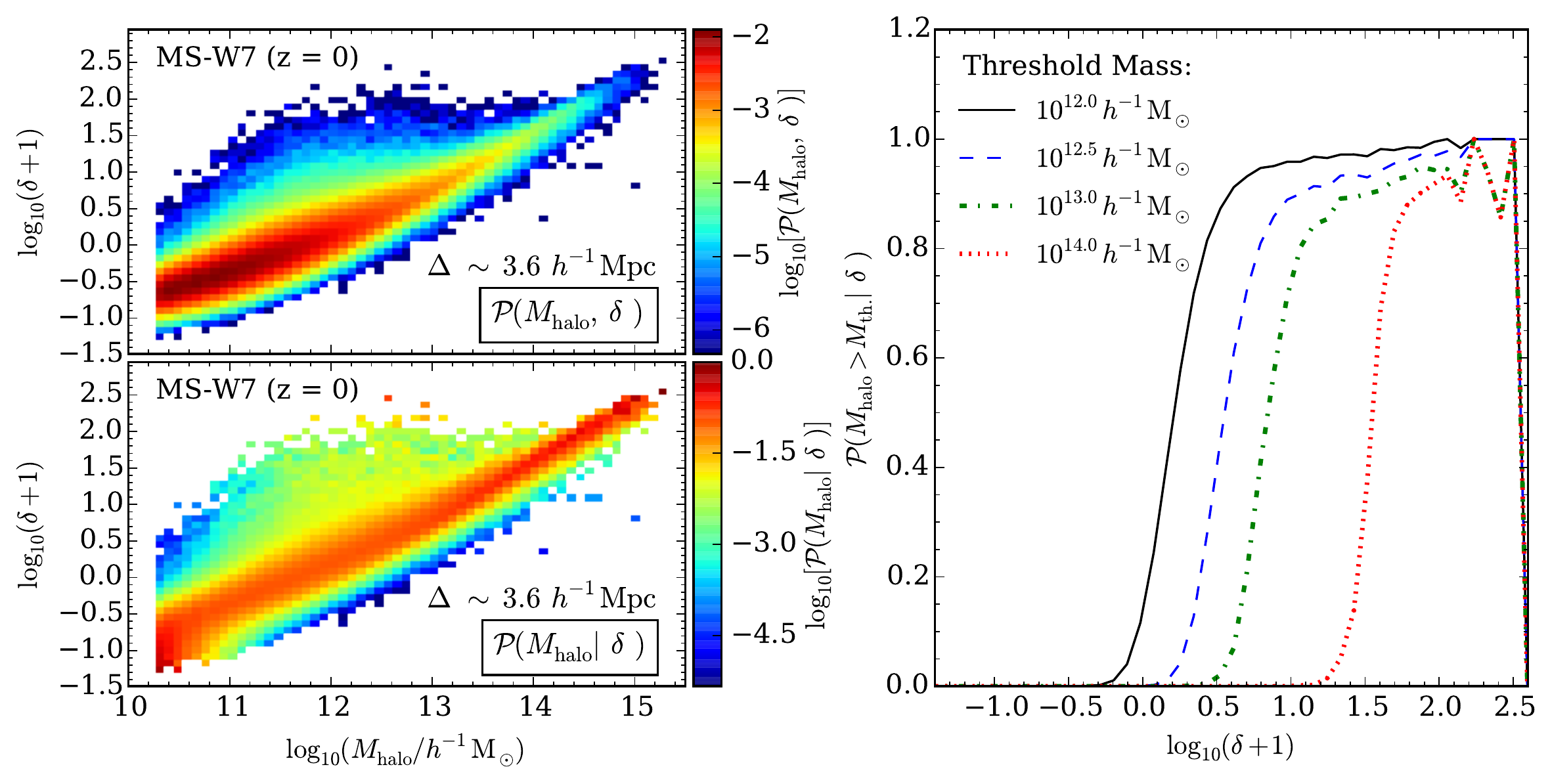}
  \caption{The upper left-hand panel shows the joint probability
  distribution $\mathcal{P}(M_{{\rm halo}},\delta)$ for the $z=0$
  snapshot of the MS-W7 simulation, where $M_{{\rm halo}}$ is the mass
  of the \emph{most massive} halo in any particular voxel. The lower
  left-hand panel shows the corresponding conditional probability
  distribution $\mathcal{P}(M_{{\rm halo}}|\delta)$, also for the
  $z=0$ snapshot of the MS-W7 simulation. This distribution shows the
  probability that, given the value for the density field, $\delta$,
  in a voxel, the most massive halo in that voxel has a mass $M_{\rm
  halo}$. The right-hand panel shows the probability
  $\mathcal{P}(M_{{\rm halo}}>M_{{\rm th.}}|\delta)$ that the most
  massive halo in a voxel has a mass greater than a threshold value,
  $M_{{\rm th}}$. Probability distributions are shown for four
  threshold masses: $10^{12.0}h^{-1}\Msol$ (black solid line),
  $10^{12.5}h^{-1}\Msol$ (blue dashed line), $10^{13.0}h^{-1}\Msol$
  (green dot-dashed line) and $10^{14.0}h^{-1}\Msol$ (red dotted
  line).}
  \label{fig:Pmd_cond_prob}
\end{figure*}

\begin{figure*}
  \centering
  \includegraphics[width=0.95\textwidth]{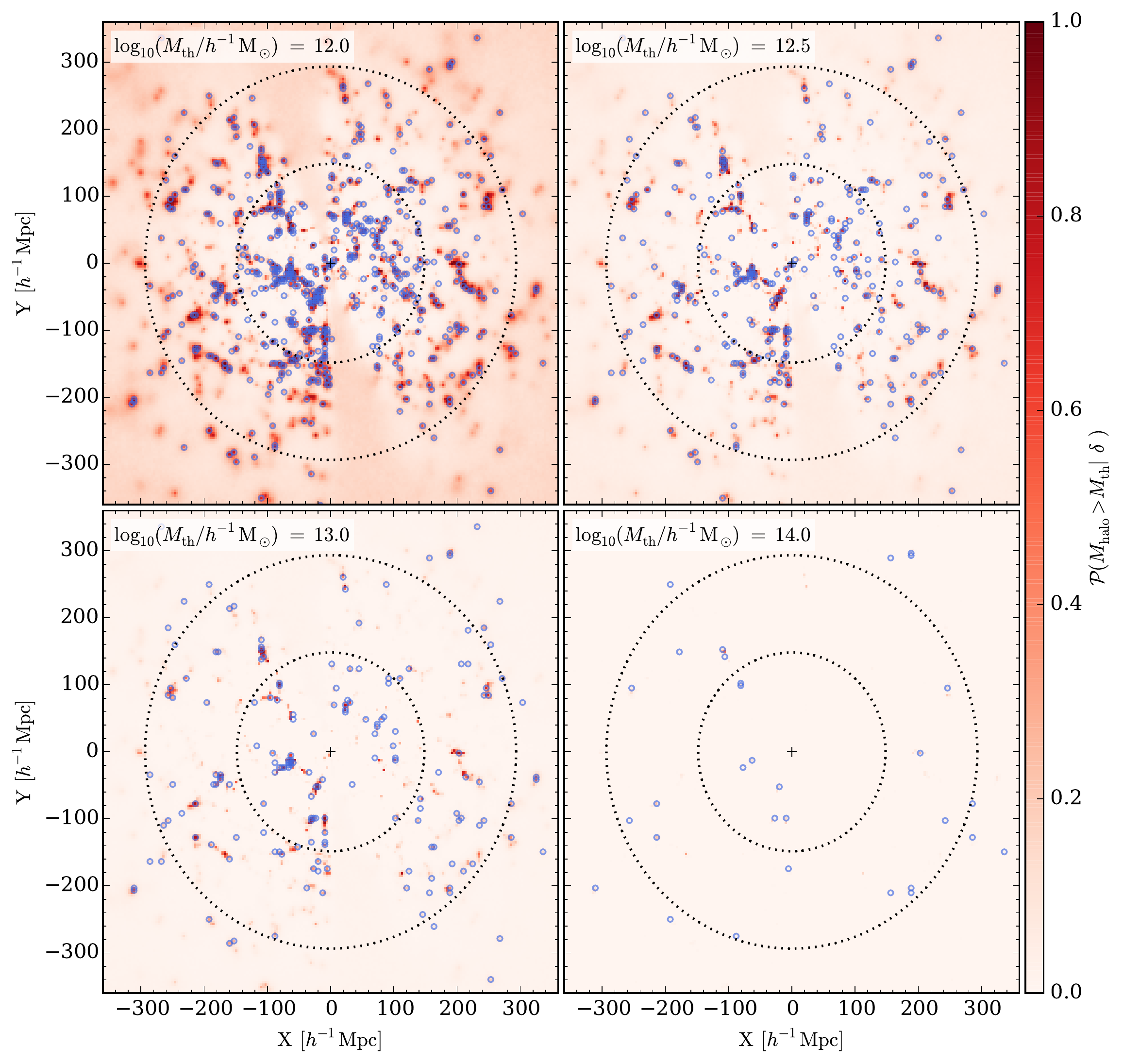}
  \caption{Slices through the \HADES{} volume at ${\rm
  Z}\sim-3h^{-1}\Msol$ showing the detection probability for four
  different mass thresholds: $10^{12.0}h^{-1}\Msol$ (top left),
  $10^{12.5}h^{-1}\Msol$ (top right), $10^{13.0}h^{-1}\Msol$ (bottom
  left) and $10^{14.0}h^{-1}\Msol$ (bottom right). Open circles show
  the positions of the voxels for which the most massive halo has a
  mass above the corresponding threshold.}
  \label{fig:halo_probs}
\end{figure*}

\begin{figure*}
  \centering
  \includegraphics[width=0.95\textwidth]{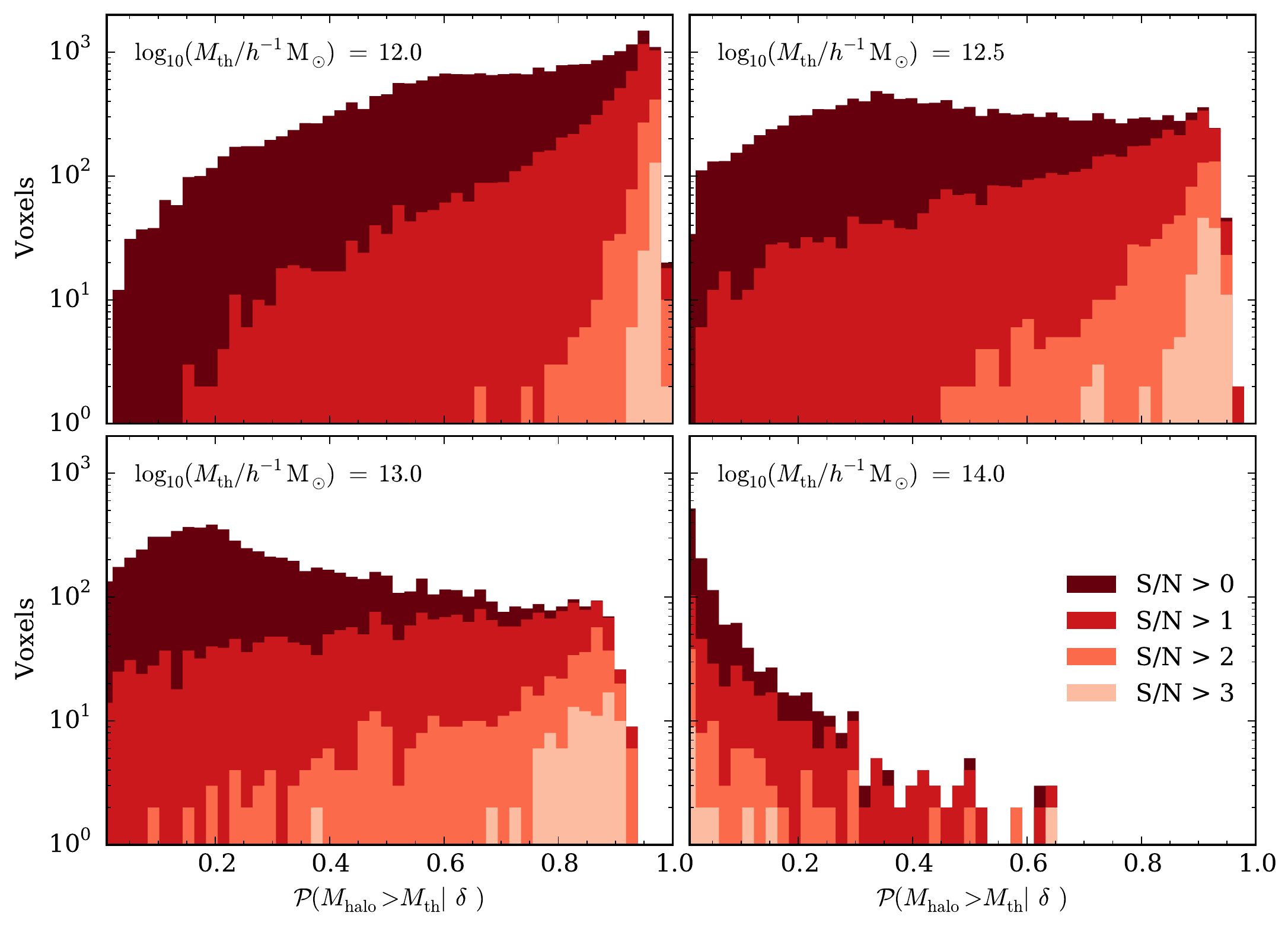}
  \caption{Distribution of detection probabilities for
  voxels whose most massive halo has a mass above $10^{12.0}h^{-1}\Msol$
  (top left), $10^{12.5}h^{-1}\Msol$ (top right), $10^{13.0}h^{-1}\Msol$ (bottom
  left) and $10^{14.0}h^{-1}\Msol$ (bottom right). The different shaded
  histograms show the impact of placing an additional cut in S/N ratio
  and considering only those voxels above a specified threshold:
  $\mathrm{S/N} > \{0,1,2,3\}$.}
  \label{fig:mass_prob_hist}
\end{figure*}

Having determined that \HADES{} is able to successfully identify the
highest S/N peaks in the density field, we now present a Bayesian
prescription that will allow us to extract information on the halo
population from the inference results. In other words, given a set of
observations, $d$, we wish to extract information on some specific
quantity, $\alpha$.

\subsection{Translating density to halo mass}
\label{sec:density_to_mass}

In Bayesian parlance, we are interested in analysing the posterior
distribution $\mathcal{P}(\alpha|d)$ and letting the data decide on
the value of $\alpha$. In our approach we can formulate the posterior
distribution $\mathcal{P}(\alpha|d)$ as a marginalisation over all
density fields, at fixed redshift, as inferred within the
\HADES{} framework:
\begin{eqnarray}
\qquad\qquad\quad\mathcal{P}(\alpha|d)&=& \int \mathrm{d}\delta \mathcal{P}(\delta,\alpha|d) \nonumber \\
&=& \int \mathrm{d}\delta \mathcal{P}(\delta|d)\, \mathcal{P}(\alpha|\delta,d)  \nonumber \\
&=& \int \mathrm{d}\delta \mathcal{P}(\delta|d)\, \mathcal{P}(\alpha|\delta)  \nonumber \\
&=& \frac{1}{N_{\mathrm{samp}}} \sum_i\mathcal{P}(\alpha|\delta_i),  
\label{eq:Blackwell_Rao_estimator}
\end{eqnarray}
where we assume conditional independence
$\mathcal{P}(\alpha|\delta,d)=\mathcal{P}(\alpha|\delta)$ once the
true density field is given, and the posterior distribution
$\mathcal{P}(\delta|d)=1/N_{\mathrm{samp}}\sum_i
\delta^D(\delta-\delta_i)$ is provided as an ensemble of data
constrained density realisations via the \HADES{} algorithm. The
chain-rule approach described in
Eq.~(\ref{eq:Blackwell_Rao_estimator}) is frequently referred to as a
Blackwell-Rao estimator. A similar approach has been implemented by
\citet{Leclercq15} to identify voids in the SDSS.

As demonstrated above, a full Bayesian quantification of unknown
quantities $\alpha$ from the observations now reduces to providing the
conditional probability distribution $\mathcal{P}(\alpha|\delta)$,
which can be simply determined from numerical simulations of structure
formation. Generally this approach can handle arbitrarily complex
problems, requiring only a determination of the corresponding
$\mathcal{P}(\alpha|\delta)$, which can be achieved via analytic or
numerical means. For the sake of this work we will exemplify this
approach to answer the question of how to find halos above a given
mass in a galaxy survey such as the 6dFGS. Specifically, the question
we wish to address is, for a voxel with a given density, $\delta$,
what is the probability that the \emph{most massive} dark matter halo
found in that voxel has a mass, $M_{{\rm halo}}$, that is larger than
a particular mass threshold, $M_{{\rm th.}}$. Given the approach of
the Blackwell-Rao estimator, as described above, this task reduces to
determining $\mathcal{P}(M_{{\rm halo}}>M_{{\rm th.}}|\delta)$, which
describes the probability of finding the most massive halo of mass
$M_{{\rm halo}}$ given a value of the density field $\delta$.

The first stage in determining $\mathcal{P}(M_{{\rm halo}}>M_{{\rm
th.}}|\delta)$ is to consider a method for translating between
density, $\delta$, and halo mass, $M_{{\rm halo}}$. This can be
achieved by tabulating the conditional probability,
\begin{equation}
\mathcal{P}(M_{{\rm halo}}|\delta)=\frac{\mathcal{P}(M_{{\rm
halo}},\delta)}{\mathcal{P}(\delta)},
\label{eq:mass_dens_conditional_prob}
\end{equation}
from the snapshot of an N-body simulation. In practice, the joint
probability, $\mathcal{P}(M_{{\rm halo}},\delta)$, can be calculated
by simply building a two dimensional histogram between $\delta$ and
$M_{{\rm halo}}$, where $M_{{\rm halo}}$ is the mass of
the most massive halo in the voxel. The joint probability distribution
is shown in the upper left-hand panel of
Fig.~\ref{fig:Pmd_cond_prob}. Here we estimate the conditional
distribution $\mathcal{P}(M_{{\rm halo}}|\delta)$ using again the
$z=0$ snapshot of the MS-W7. We estimate the density field for the
simulation by binning the dark matter particles into a grid of $139^3$
voxels. Given the size of the simulation box, $500\hMpc$ on a side,
this gives a resolution of $\sim3.6\hMpc$, approximately identical to
the resolution used in our \HADES{} inference analysis. Note that we
do not use the density field calculated according to the \HADES{}
volume as we do not want to bias the conditional probability by
introducing repeated structures. The conditional probability
distribution, shown in the lower left-hand panel of
Fig.~\ref{fig:Pmd_cond_prob}, is the conditional probability that the
most massive halo in a $3.6\,\hMpc$ voxel with a given density,
$\delta$, will have a mass of $M_{{\rm halo}}$. The distribution shows
a clear, monotonic relation that we can use to translate between the
density of a voxel and the mass of the most massive halo within that
volume element. In reality the distribution $\mathcal{P}(M_{{\rm
halo}}|\delta)$ will have an additional redshift dependence, which
could be modelled by computing $\mathcal{P}(M_{{\rm halo}}|\delta)$
for each snapshot of the simulation and interpolating between the
distributions. However, given that the 6dFGS is a very shallow survey,
with median redshift $z_{\mathrm{med}}\sim0.05$, for the purposes of
demonstrating our methodology we can simply approximate the matter
density field through the 6dFGS mock using the $z=0$ snapshot. We have
examined the distribution $\mathcal{P}(M_{{\rm halo}}>M_{{\rm
th.}}|\delta)$ from the MS-W7 snapshots for redshifts up to $z\sim0.2$
(the approximate radial extent of the 6dFGS mock) and find negligible
evolution of the distribution away from the $z=0$ distribution.

From $\mathcal{P}(M_{{\rm halo}}|\delta)$ we can make an estimate for
$\mathcal{P}(M_{{\rm halo}}>M_{{\rm th}}|\delta)$ by marginalising
over all halo masses above the threshold halo mass, $M_{{\rm
th}}$. The right-hand panel of Fig.~\ref{fig:Pmd_cond_prob} shows
estimates for $\mathcal{P}(M_{{\rm halo}}>M_{{\rm th.}}|\delta)$, at
$z\sim0$, for four different mass thresholds: $M_{{\rm
th.}}=10^{12.0}h^{-1}\Msol$, $10^{12.5}h^{-1}\Msol$, $10^{13.0}h^{-1}\Msol$ and
$10^{14.0}h^{-1}\Msol$. For each mass threshold, the detection probability
for a halo undergoes quite a sharp transition as a function of
density. Furthermore, the transition of the probability from zero to
one occurs at higher densities for larger mass thresholds. We note
that the detection probability drops back down to zero at
$\log_{10}(\delta+1)\sim2.5$ due to the limited volume of the MS-W7
simulation. However, for a larger volume simulation, above
$\log_{10}(\delta+1)\sim2.5$ the detection probability would remain
constant at unity.

\begin{figure}
  \centering
  \includegraphics[width=0.45\textwidth]{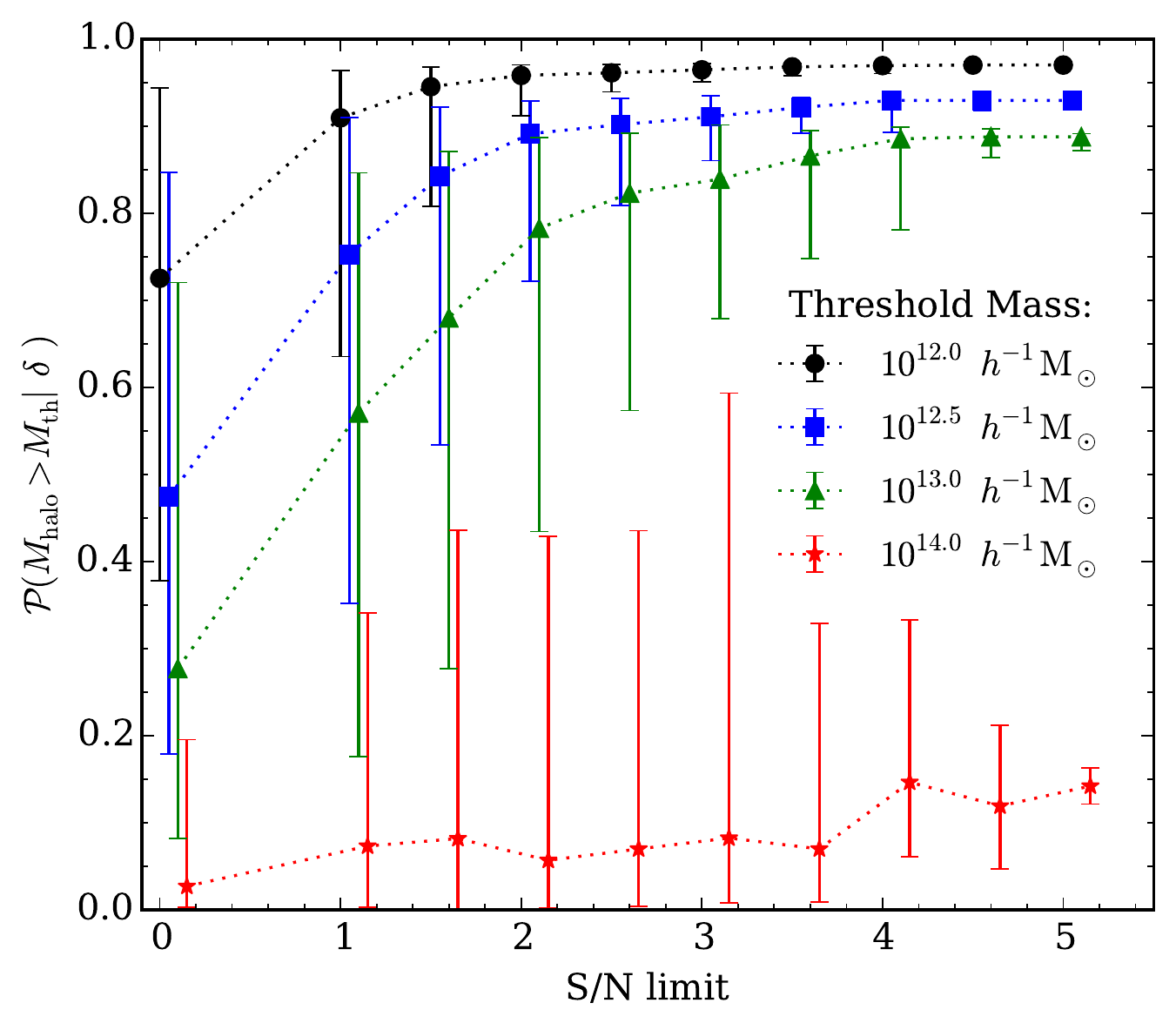}
  \caption{Change in detection probability as a function
  of signal-to-noise limit. The filled symbols show the median
  detection probability for those voxels that have a S/N above the
  corresponding limit and whose most massive halo has a mass above
  $10^{12.0}h^{-1}\Msol$ (circles), $10^{12.5}h^{-1}\Msol$ (squares),
  $10^{13.0}h^{-1}\Msol$ (triangles) and $10^{14.0}h^{-1}\Msol$ (stars). The error
  bars show the $\mathrm{10^{th}}$ and $\mathrm{90^{th}}$
  percentiles.}
  \label{fig:median_prob}
\end{figure}

\subsection{Detection probability maps}
\label{sec:detection_maps}

Using this result we are therefore able to build maps of the detection
probability for halos above specific threshold masses given some
galaxy observation, $d$. These maps are built by using the
Blackwell-Rao approach, as described in
Eq.~(\ref{eq:Blackwell_Rao_estimator}), and simply marginalising over
all data constrained realisations of the density field obtained via
the \HADES{}, using the distributions in the right-hand panel of
Fig.~\ref{fig:Pmd_cond_prob} to assign a weight to each voxel. In
Fig.~\ref{fig:halo_probs} we show maps for the halo detection
probabilities for the four different mass thresholds:
$\log_{10}(M_{{\rm th.}}/h^{-1}{\rm
M_{\odot}})=\{12.0,12.5,13.0,14.0\}$. Given the mock catalogue and
knowledge of the underlying halos, we can determine the mass of the
most massive halo in each \HADES{} voxel. (Note that this information
is stored when we build the mock catalogue, before any geometrical,
photometric or completeness limits are applied.) On top of the
detection maps we indicate with blue circles those voxels whose most
massive halo is above the specified threshold. We stress,
however, that these detection maps are not only reconstructions of the halo
distribution, but instead, for any position $\bar{x}$, quantify our
belief that there exists a halo above a given mass threshold located
at that point. This provides a natural quantification of detection
uncertainties in the survey.

\begin{figure*}
  \centering
  \includegraphics[width=0.98\textwidth]{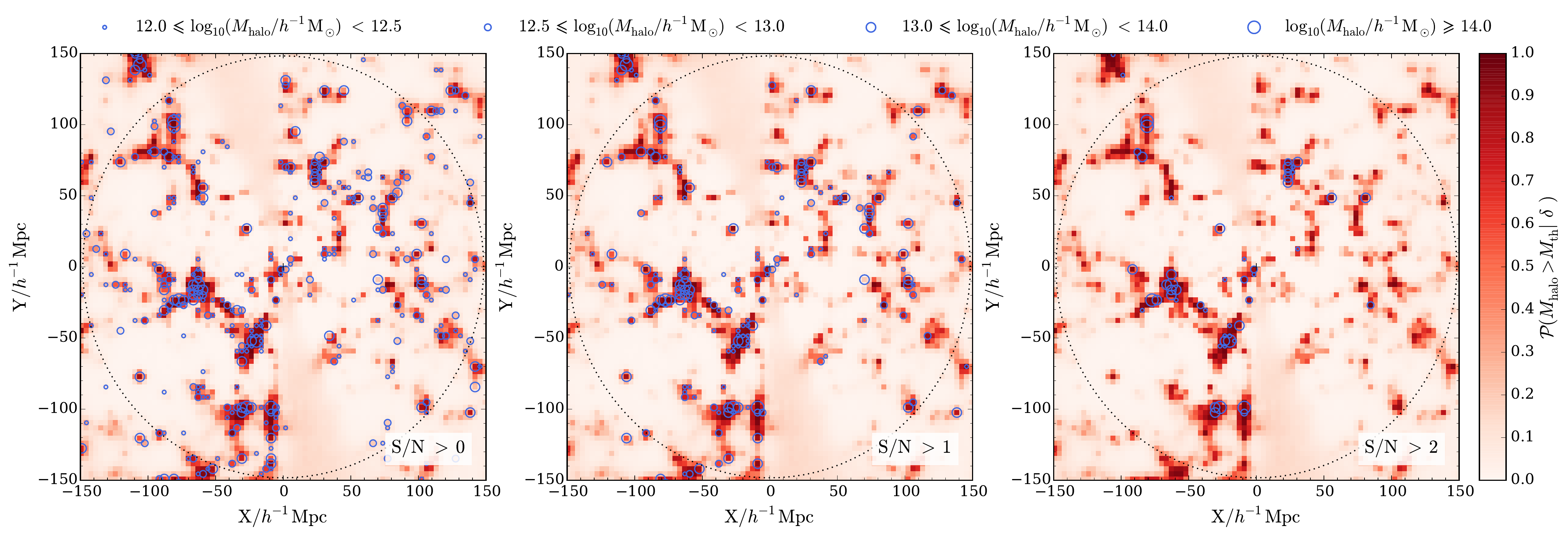}
  \caption{Three zoom in slices of the detection probability map for
  $M_{\mathrm{th}}=10^{12}h^{-1}\Msol$, showing the region within the median
  redshift of the mock catalogue (as indicated by the dotted
  circle). The blue circles show the positions of the voxels with a
  signal-to-noise ratio (S/N) above the specified threshold and whose
  most massive halo is within the particular mass bin. The S/N ratio
  thresholds are: $\mathrm{S/N} > 0$ (left-hand panel), $\mathrm{S/N}
  > 1$ (middle panel) and $\mathrm{S/N} > 2$ (right-hand panel).}
  \label{fig:mass_prob_zooms}
\end{figure*}

For the three lowest mass thresholds it can be seen that the detection
probability for halos of respective masses is fairly high close to the
observer where the survey generally exhibits high signal-to-noise
ratios. As can be seen, many halos are correctly identified by the
relative peaks in the detection probability. With increasing distance
from the observer the detection of respective halo populations becomes
increasingly uncertain. This is because, due to flux limitations of
the survey, we only observe the brighter objects that are typically
hosted by more massive halos at larger distances. Dim objects
corresponding to less massive halos have a vanishing probability of
being detected by the flux limited survey. As can be seen in
Fig.~\ref{fig:halo_probs}, the respective panels correctly reflect
this behaviour.

  For the $M_{\mathrm{th.}}=10^{14}h^{-1}\Msol$ mass threshold,
  however, we see, on first inspection, very few
  detection peaks, with several halos appearing not to have a
  corresponding peak in the detection probability map. We
  see that, given our observational dataset, several of these
  mis-detections occur in noise-dominated regions, where we have only
  a handful of galaxies. If, however, we were to artificially boost the
  detection probabilities in the map, we would see that many of the
  halos do indeed correspond to relative peaks in the probability and
  that these peaks simply have a lower amplitude compared to the peaks
  in the detection maps for the other mass thresholds. This is due to
  our cosmological model and our prior belief of finding halos above a
  particular mass, which is encoded in the matter power spectrum. The
  $\mathrm{\Lambda}$CDM cosmological model predicts that in a given
  volume, such as that of the MS-W7 simulation, we should expect to
  find relatively few high density peaks compared to low density peaks
  and so would expect to find fewer high mass halos compared to lower
  mass halos. Suppose therefore we were to bet on finding a halo above
  $10^{14}h^{-1}\Msol$ at a particular position. Given our
  cosmological model, for noise-dominated regions we would be less
  confident and would not bet as highly on finding a halo above a
  higher mass threshold. As such, given the observational dataset, our
  halo detection methodology assigns a non-zero detection probability,
  but is conservative due to our physical expectation that we are
  generally less likely to find an extreme event. In a similar
  fashion, our methodology encodes the fact that we are more likely to
  detect a lower mass halo and so assigns a higher detection
  probability for lower mass thresholds.

  There are several factors which could act to further smooth the
  amplitude of the detection probability peaks. Firstly the fact that
  we have fewer density high density peaks leads to the
  $\mathcal{P}(M_{{\rm halo}}|\delta)$ conditional probability, shown
  in the lower left-hand panel of Fig.~\ref{fig:Pmd_cond_prob},
  becoming noisier towards larger densities and halo masses. This
  increases the width of $\mathcal{P}(M_{{\rm halo}}|\delta)$, thus
  causing a particular density amplitude to correspond to a range of
  halo masses. As a result, more massive halos could potentially be
  mistaken for lower mass objects. Using a simulation with larger
  cosmological volume would help prevent this. Secondly, our modelling
  of phenomena such as galaxy bias could lead to a systematic offset
  between the density amplitudes in the simulation and the density
  amplitudes recovered by \HADES{}. In this work we have assumed a
  fixed bias of $b=1.2$. The impact of galaxy bias could in future
  work be examined by reproducing the \HADES{} inference analysis
  using a range of different bias values, though the ability of
  \HADES{} to infer luminosity-dependent galaxy bias is also currently
  being tested. Finally, another important factor is redshift-space
  effects. The \HADES{} reconstructions correspond to the
  redshift-space density field, whilst the calculated
  $\mathcal{P}(M_{{\rm halo}}|\delta)$ corresponds to the real-space
  density field of the N-body simulation. Redshift-space effects, such
  as fingers-of-god effects, act to smooth out real-space density
  peaks, especially density peaks. As such, this could again lead to a
  high mass halo being mistaken as a lower mass halo. The impact of
  redshift-space distortions in \HADES{} is still being investigated
  \citep[see ][]{Jasche12} and will be considered in future work.

\subsection{Recovery of individual clusters}
\label{sec:recovery_of_clusters}

To begin to quantify the success of the detection of halos we examine
the distribution of probabilities for those voxels whose most massive
halo is above the different mass thresholds. We plot these
distributions, for each of the four mass thresholds, in
Fig.~\ref{fig:mass_prob_hist}. When considering all such voxels with a
signal-to-noise ratio greater than zero, we see that, with the
exception of the $10^{12}h^{-1}\Msol$ mass threshold, every
distribution peaks at low probabilities. This is because,
as discussed in the previous section, in noisier regions with lower
signal-to-noise we have less confidence of detecting higher mass
halos. We would therefore expect such voxels to be poorly constrained
by \HADES{}, leading to a reduced detection probability. 

We show in Fig.~\ref{fig:mass_prob_hist}, how the
distribution of detection probabilities changes as we restrict
ourselves to voxels with higher S/N ratios: $\mathrm{S/N} > 1$,
$\mathrm{S/N} > 2$ and $\mathrm{S/N} > 3$. As the S/N limit is
increased the peak of the distribution shifts towards higher detection
probabilities. In Fig.~\ref{fig:median_prob} we plot the change in the
median detection probability as a function of S/N ratio. The increase
in the median probability with increasing S/N ratio reflects our
confidence in detecting a higher mass halo. For the highest mass
threshold, $M_{\mathrm{th.}}=10^{14}h^{-1}\Msol$, we see a
consistently low detection probability, as we have discussed
previously. Note however that this mass threshold still displays a
median probability that increases with increasing S/N, reflecting our
increasing confidence of detecting a halo with mass above
$10^{14}h^{-1}\Msol$ in highly constrained voxels.

As such, expressing the success of our detection methodology becomes a
function of S/N. We demonstrate this visually in
Fig.~\ref{fig:mass_prob_zooms}, where we zoom in on
$M_{\mathrm{th.}}=10^{12}h^{-1}\Msol$ probability map for the region
within the median redshift of the mock catalogue. In the three
consecutive panels we overlay the positions of voxels with a S/N above
a particular limiting value and where the most massive halo in that
voxel is within a particular mass range. For the $\mathrm{S/N} > 0$
panel we can see that there are several mis-detections, particularly
for lower-mass halos. However, as we increase the S/N ratio we can see
that the number of mis-detections decreases and the positions of the
halos correlate well with large peaks in the detection probability.

Finally, we stress that this analysis serves as a proof of concept,
where we have used a simple measurement task to demonstrate the
feasibility of our Bayesian halo detection approach, as outlined
above. However, the method only relies on the conditional distribution
$\mathcal{P}(\alpha|\delta)$ of some quantity $\alpha$ given a density
field $\delta$ (at a redshift $z$), which can either be generated via
analytic calculations or extracted from simulations as described
here. For this reason the proposed Bayesian detection methodology is a
flexible and versatile approach that can be arbitrarily increased in
complexity to test various quantities and features of the cosmic
large-scale structure in cosmological datasets. The excellent
agreement between the peaks in our detection probability maps and the
positions of high S/N halos indicates that this methodology could be
used in the construction of an accurate catalogue of probabilistic
cluster candidates, though a resolution finer than $3.6\hMpc$ would
likely be required.


\section{Summary \& Conclusions}
\label{sec:conclusions}
We present a novel Bayesian methodology for inferring various
properties of the cosmic large-scale structure. Specifically, we focus
on determining the detection probability of halos with masses above
different thresholds in cosmological observations, which may be
subject to stochastic and systematic uncertainties. Our approach
relies on the previously developed \HADES{} algorithm, designed to
infer the smooth matter density field of the cosmic large-scale
structure in the non-linear regime, and the Blackwell-Rao
Estimator, which we use to relate density field amplitudes to halo
properties. In this work we present a proof-of-concept of our
methodology by applying it to a realistic galaxy mock catalogue for
which the halo positions and membership are already known.

We construct a realistic galaxy mock catalogue by populating the halos
of a cosmological N-body simulation with galaxies from a
semi-analytical galaxy formation model. The mock catalogue emulates
the \band{K}-band selected catalogue of the 6dFGS final data release
(DR3). We apply the \HADES{} algorithm to the mock catalogue in four
parallel Markov chains to generate a total of 20,000 realisations of
the matter density field through approximately $0.5\,h^{-3}{\rm
Gpc}^{3}$ of the volume of the mock catalogue, sampled at a resolution
of approximately $3.6\hMpc$. Examination of recovery of the matter
power spectrum suggests that the Markov chains converge within
approximately 2000 samples. As a conservative measure, however, we
remove the first 5000 samples from each chain to allow for burn-in,
which leaves us with a total of 20,000 independent \HADES{}
realisations of the density field.

We present the ensemble mean and variance of the density field
recovered by \HADES{}. Despite the Gaussian nature of each individual
sample, the ensemble mean density field is distinctly non-Gaussian,
with large-scale structures such as galaxy clusters and voids, which
constitute high signal-to-noise features, clearly identifiable out to
twice the median redshift of the mock survey. To quantify the success
of the recovery of structures by \HADES{} we consider the correlation
between the \HADES{} density field and the MS-W7 density field, as
estimated within the \HADES{} volume. Examining the Pearson
correlation as a function of \HADES{} ensemble density we find a high
correlation in the highest and lowest density bins. This result
indicates that \HADES{} is successfully recovering high
signal-to-noise regions, such as clusters and voids.

Finally we present a Bayesian prescription to address the problem of
extracting information for the halo population from a set of
observations from a galaxy survey. Specifically, we use a
Blackwell-Rao estimator to address the question, given a value for the
density field, $\delta$, over a volume element at redshift, $z$, what
is the probability, $\mathcal{P}(M_{{\rm halo}}>M_{{\rm
    th.}}|\delta)$, that the most massive halo within that volume has
a mass, $M_{{\rm halo}}$, greater than some threshold value, $M_{{\rm
    th.}}$. A cosmological simulation can be used to construct the
conditional probability $\mathcal{P}(M_{{\rm halo}}|\delta)$ for the
mass of the most massive halo in a volume element. By marginalising
over all \HADES{} realisations and using the density amplitude to
weight each voxel according to $\mathcal{P}(M_{{\rm halo}}>M_{{\rm
    th}}|\delta)$, we can construct maps of the detection probability
for halos above selected threshold masses. For each mass threshold
considered, the relative peaks in the detection probability correspond
quite well to the positions of halos with masses above the
threshold. However, for the highest mass threshold of
$10^{14}h^{-1}\Msol$ the peaks in the detection probability have lower
amplitude, which leads to an increasing number of apparent
mis-detections. This is due to our cosmological model, which predicts
that we should expect to find relatively few high mass halos compared
to lower mass halos. As such, our methodology encodes this expectation
and reflects our reduced confidence of detecting very massive halos,
especially in regions of low signal-to-noise. This means, for example,
that with increasing distance from the observer the probability of
detection of more massive halo populations becomes increasingly
uncertain. We find therefore that the success of the detection method
is a function of the S/N ratio. For the three lowest mass thresholds,
halos in voxels with $\mathrm{S/N}>1$ are typically detected with a
probability greater than 0.5, whilst halos in voxels with
$\mathrm{S/N}>2$ are typically detected with a probability in excess
of 0.8.

Our Bayesian description provides a statistically thorough approach to
quantify the detection probability and corresponding uncertainties for
halos above a given mass threshold. Following this proof-of-concept we
plan to, in future work, apply \HADES{} and our halo detection
prescription to the actual 6dFGS observational data. Beyond this our
methodology can be applied to mock catalogues and actual observations
of deeper spectroscopic surveys, in order to demonstrate the ability
of our methodology to detect halos out at higher redshifts.  We stress
however that our methodology is versatile and can be applied to a wide
variety of datasets, including deep catalogues of galaxies with
photometric redshifts (thanks to the photometric redshift sampling
that is possible with \HADES{}). Therefore, we aim in future work to
additionally apply the methodology to photometric datasets. As such,
the Bayesian methodology that we have presented offers a promising
approach for the analysis of ongoing and future large-scale structure
surveys.

\section*{Acknowledgements}
We thank the anonymous referee for many thorough and constructive
comments. In addition, we also thank Sreekumar Thaithara Balan, Boris
Leistedt, Michelle Lochner and Hiranya Peiris for several productive
and insightful discussions and suggestions. FBA acknowledges the
support of the Royal Society for a University Research Fellowship. OL
acknowledges support from a European Research Council Advanced Grant
FP7/291329. BDW acknowledges support from NSF grants AST 07-08849 and
AST 09-08693 ARRA, and a Chaire d'Excellence from the Agence Nationale
de Recherche. This research was supported by the DFG cluster of
excellence "Origin and Structure of the Universe"
(www.universe-cluster.de).

\bibliographystyle{mn2e_trunc8}
\bibliography{aimerson}

\end{document}